 \documentclass[%
  aip,jcp,
 amsmath,amssymb,
  reprint,%
 ]{revtex4-1}

\usepackage{graphicx}% Include figure files
\usepackage{dcolumn}% Align table columns on decimal point
\usepackage{bm}% bold math
\usepackage[utf8]{inputenc}
\usepackage[T1]{fontenc}
\usepackage{mathptmx}

\usepackage{epstopdf}
\usepackage[usenames,dvipsnames]{xcolor}
\usepackage[caption=false]{subfig}
\usepackage{booktabs}

\usepackage{soul}
\usepackage{color}

\begin{document}

\preprint{AIP/123-QED}

\title[Unsupervised learning for local structure detection in colloidal systems]{Unsupervised learning for local structure detection in colloidal systems}
% Force line breaks with \\

\author{Emanuele Boattini}
 \email{e.boattini@uu.nl.}
\author{Marjolein Dijkstra}%

\author{Laura Filion}
\affiliation{ 
Soft Condensed Matter, Debye Institute for Nanomaterials Science, Utrecht University, Utrecht, The Netherlands%\\This line break forced with \textbackslash\textbackslash
}%

\date{\today}% It is always \today, today,

\begin{abstract}
We introduce a simple, fast, and easy to implement unsupervised learning algorithm for detecting different local environments on a single-particle level in colloidal systems. In this algorithm, we use a vector of standard bond-orientational order parameters to describe the local environment of each particle. We then use a neural-network-based autoencoder combined with Gaussian mixture models in order to autonomously group together similar environments. We test the performance of the method on snapshots of a wide variety of colloidal systems obtained via computer simulations, ranging from simple isotropically interacting systems, to binary mixtures, and even anisotropic hard cubes. Additionally, we look at a variety of common self-assembled situations such as fluid-crystal and crystal-crystal coexistences, grain boundaries, and nucleation. In all cases, we are able to identify the relevant local environments to a similar precision as "standard", manually-tuned and system-specific, order parameters. In addition to classifying such environments, we also use the trained autoencoder in order to determine the most relevant bond orientational order parameters in the systems analyzed. 
\end{abstract}

\maketitle

%%%MAIN TEXT%%%%
\section{Introduction}
An important challenge in the study of colloidal self-assembly is the detection of self-assembled products in the system. When the basic phases that the system forms are well characterized, we can design order parameters that detect, on a single particle level, which of the expected phases a particle is in. This strategy has been extensively used in studies ranging from crystal nucleation and growth \cite{auer2001prediction} \cite{gasser2001real,hermes2011nucleation,sanz2008out}, to crystal melting, and even grain boundary dynamics \cite{gb,skinner2010grain}. Over the years, a number of different routes have been taken to characterise such local order, including e.g. order parameters based on bond-orientational order \cite{steinhardt1983bond,ten1995numerical,rein1996numerical,Lechner2008}, common neighbor analysis (CNA)\cite{honeycutt1987molecular, faken1994systematic}, and templating \cite{duff2011polymorph}. Moreover, developments in the last few years have started to combine these local descriptions with supervised machine learning techniques in order to recognize specific crystal structures \cite{geiger2013neural, dietz2017machine, Boattini2018}. These strategies, however, only work for systems where we know a priori which phases we expect to find.

In many cases, when exploring the self-assembly of new systems, the exact final structure and its characteristics are unknown, complicating the selection of an ideal order parameter. In this context, unsupervised machine learning techniques, which excel in autonomously finding patterns in large data sets, offer a promising route for detecting self-assembled structures. 
In a recent paper \cite{reinhart2017machine}, Reinhart {\it et al.} made an attempt to use unsupervised machine learning in order to identify different crystal structures. 
In their work, they described the local environment of the particles with the adjacency criterion from adaptive CNA and combined it with the diffusion map technique for dimensionality reduction in order to distinguish different, frequently occurring, structures. Although very successful, this method turned out to be very computationally demanding. In a subsequent article \cite{reinhart2018automated}, a faster way of comparing local neighborhoods was introduced, based on their relative graphlet frequencies. This reduced the computational cost of the algorithm by four orders of magnitude.

A different approach was followed by Spellings and Glotzer in Ref. \onlinecite{spellings2018machine}, where they used a combination of unsupervised and supervised learning techniques in order to identify the overall crystal structures of bulk self-assembled systems (i.e. systems of which the majority had self-assembled into the same phase).

In this work, we present a new avenue to detect self-assembly products, and introduce an unsupervised machine learning algorithm based on bond-orientational order parameters combined with neural-network-based autoencoders\cite{Rumelhart1986, Kramer1991, Scholz2002, Goodfellow, Bishop} and Gaussian mixture models \cite{EM, BIC, Baudry2010}. Autoencoders are a standard technique for nonlinear dimensionality reduction, while mixture models are probabilistic models for identifying distinct clusters within a data set. Using these methods, our algorithm can autonomously classify particles in different groups based on their local order, making it easy to detect any self-assembly product in the system.

In contrast to Ref. \onlinecite{spellings2018machine}, the goal here is to identify local environments on a single-particle level - meaning that this method can be used to study processes like nucleation, grain boundary characteristics, and coexistences. This algorithm has been designed to be computationally fast, easily scalable to very large data sets, and extremely easy to implement. To test its performance, we examine a number of different colloidal systems, ranging from spheres, to binary mixtures, to anisotropic particles. Moreover, in addition to simply classifying local environments, we also explore whether the unsupervised learning techniques employed can help us identify the distinguishing features of the different particle environments found in the system.

\section{Methods}

In this section, we describe in detail the algorithm we use to classify local environments. We start by summarizing the main steps of our approach, and then follow with a detailed description of each step in separate subsections. 

The overall process consists of three steps. First, we require a method to capture the local environment of each particle in a set of local order parameters. For this, we make use of bond-orientational order parameters. This set of order parameters is in general high-dimensional, and may contain significant amounts of redundant and irrelevant information. In order to extract the most relevant information, the second step of our approach makes use of a dimensionality reduction technique, namely a neural-network-based autoencoder. Once trained, the autoencoder projects the original (high-dimensional) input vectors onto a lower-dimensional subspace encoding the features with the largest variations in the input data. Ideally, in this subspace, particles with similar local environments are grouped together. Finally, we apply a clustering algorithm (Gaussian mixture models) in order to identify the distinct clusters of local environments in this lower-dimensional subspace.

\subsection{Bond order parameters}
\label{sec:BOP}

To characterize the local environment of each particle, we use the averaged bond order parameters (BOPs) introduced by Lechner and Dellago \cite{steinhardt1983bond, Lechner2008}. First, we define for any given particle $i$ the complex quantities
\begin{equation}
\label{qlm}
q_{lm}(i) = \frac{1}{N_b(i)}\sum_{j \in \mathcal{N}_b(i)} Y^m_l(\mathbf{r}_{ij}),
\end{equation}
where $Y^m_l(\mathbf{r_{ij}})$ are the spherical harmonics of order $l$, with $m$ an integer that runs from $m=-l$ to $m=+l$. Additionally, $\mathbf{r}_{ij}$ is the vector from particle $i$ to particle $j$, and $\mathcal{N}_b(i)$ is the set of nearest neighbors of particle $i$, which we will define later.  Note that $\mathcal{N}_b(i)$ contains $N_b(i)$ particles.  Then, we can define an average $\bar{q}_{lm}(i)$ as
\begin{equation}
\label{avqlm}
\bar{q}_{lm}(i) = \frac{1}{{N}_b(i) +1}\sum_{k\in \{i, \mathcal{N}_b(i)\}} q_{lm}(k),
\end{equation}
where the sum runs over all nearest neighbors of particle $i$ as well as particle $i$ itself. Averaging over the nearest neighbor values of $q_{lm}$ results effectively in also taking next-nearest neighbors into account. 
%Finally, we define rotationally invariant quadratic and cubic order parameters as
Finally, we define rotationally invariant BOPs as
\begin{equation}
\label{avql}
\bar{q}_l(i) = \sqrt{\frac{4\pi}{2l+1}\sum_{m=-l}^{l}|\bar{q}_{lm}(i)|^2},
\end{equation}
%\begin{equation}
%\label{avwl}
%\bar{w}_l(i) = \frac{\sum\limits_{m_1+m_2+m_3=0}^{}\left(\begin{matrix}l & l & l\\ m_1 & m_2 & m_3\end{matrix}\right)
%\%bar{q}_{lm_1}(i)\bar{q}_{lm_2}(i)\bar{q}_{lm_3}(i)}{\left(\sum_{m=-l}^{l}|\bar{q}_{lm}(i)|^2\right)^{3/2}},
%\end{equation}
%where the term in parentheses in Eq. \ref{avwl} is the Wigner $3j$ symbol. The quantities in Eqs. \ref{avql} and \ref{avwl} are real, translationally and rotationally invariant, and, depending on the choice of $l$, are sensitive to different crystal symmetries.  %%to see
which, depending on the choice of $l$, are sensitive to different crystal symmetries.

The optimal set of BOPs to be considered strongly depends on the structures one wishes to distinguish. Since our method is meant to be applied to systems for which such prior knowledge is missing, in order to describe the local environment of one particle, we evaluate several $\bar{q}_l$ with $l$ ranging from $1$ to $8$. Note that in principle, one could consider a larger (or smaller) range of $l$. For all cases examined in this paper, however, we found 8 to be sufficient. Therefore, when considering one component systems, our description of the local environment of particle $i$ is encoded into an 8-dimensional vector
\begin{equation}
\label{input1}
\mathbf{Q}(i) = (\{\bar{q}_l(i)\}),
\end{equation}  
with $l\in[1,8]$. When considering binary mixtures, i.e. systems with two species of particles, the same BOPs are evaluated both considering all the nearest neighbors of the reference particle (regardless of particles' species), and considering only the nearest neighbors of the same species as the reference particle. Hence, for binary mixtures, our description of the local environment of particle $i$ is encoded into a 16-dimensional vector
\begin{equation}
\label{input2}
\mathbf{Q}(i) = (\{\bar{q}_l(i)\}, \{\bar{q}^{ss}_l(i)\})
\end{equation}
where $s$ indicates the particles' species. Here, $\{\bar{q}_l\}$ represent the set of BOPs evaluated considering all the nearest neighbors of particle $i$, while the set $\{\bar{q}^{ss}_l\}$ is evaluated considering only the nearest neighbors  of the same species $s$ as particle $i$. 
%To describe the local environment of one particle, we evaluate several of such bond order parameters with $l$ ranging from $1$ to $8$. Note that the cubic terms, $\bar{w}_l$, are non-vanishing only for even values of $l$. Hence, when considering one component systems, our description of the local environment of particle $i$ is encoded into a 12-dimensional vector
%\begin{equation}
%\label{input1}
%\mathbf{Q}(i) = (\{\bar{q}_l(i)\}, \{\bar{w}_{l'}(i)\}),
%\end{equation}  
%with $l\in[1,8]$ and $l'$ consisting of only even values of $l$.

%When considering binary mixtures, i.e systems with two species of particles, the same bond order parameters are evaluated both considering all the nearest neighbors of the reference particle (regardless of particles' species), and considering only the nearest neighbors of the same species as the reference particle. Hence, for binary mixtures, our description of the local environment of particle $i$ is encoded into a 24-dimensional vector
%\begin{equation}
%\label{input2}
%\mathbf{Q}(i) = (\{\bar{q}_l(i)\}, \{\bar{w}_{l'}(i)\}, \{\bar{q}^{aa(bb)}_l(i)\}, \{\bar{w}^{aa(bb)}_{l'}(i)\})
%\end{equation}
%for particles of species $a$($b$). Here $\bar{q}_l$ and $\bar{w}_l$ represent the averaged bond order parameters evaluated considering all the nearest neighbors of the particles, while $\bar{q}^{aa(bb)}_l$ and $\bar{w}^{aa(bb)}_l$ are evaluated considering only the nearest neighbors  of species $a$($b$) of particles of species $a$($b$).

Thus far, we have not discussed the definition of a nearest neighbor, as used in the definition of the BOPs. There are a number of different avenues for identifying nearest neighbors. The simplest method relies on using a fixed cutoff radius $r_c$, such that all particles closer than this distance are considered nearest neighbors. Ideally, this cutoff radius is chosen as the distance at which the radial distribution function has its first minimum. This method has the advantage that it is computationally very cheap and it is symmetric, i.e. $i$ is a neighbor of $j$ if and only if $j$ is a neighbor of $i$. However, $r_c$ is system and density dependent, so that it has to be tuned for every particular case requiring prior knowledge of the system under study. Additionally, the cutoff is defined for the entire system and, as such, is not an optimal choice for systems with large density gradients or interfaces, such as can occur in nucleation studies.

Another standard method for determining nearest neighbors is the Voronoi construction \cite{Voronoi}, which has the advantage that it is parameter free. However, it is also relatively computationally expensive, and in this work we have instead opted to make use of a recently introduced alternative parameter-free nearest-neighbor criterion, called SANN (solid angle nearest neighbor) \cite{SANN}. In this approach, an effective individual cutoff is found for every particle in the system based on its local environment. %The algorithm can briefly be described as follows. 
%First, the particles $\{j\}$ surrounding $i$ are ordered such that $r_{i,j}\le r_{i,j+1}$. Then, starting with the particle closest to $i$, a solid angle $\theta_{i,j}$ is associated to every potential neighbor $j$. Finally, SANN defines the neighborhood of particle $i$ as consisting of the nearest (i.e. closest) $m$ particles $\{j\}$ such that the sum of their solid angles equals at least $4\pi$. For a complete description of the algorithm see Ref. \onlinecite{SANN}.
This method is not inherently symmetric, i.e. $j$ might be a neighbor of $i$ while $i$ is not a neighbor of $j$. However, symmetry can be enforced by either adding $j$ to the neighbors of $i$ or removing $i$ from the neighbors of $j$. In this study, we applied the latter solution. The computational cost of SANN only slightly exceeds that of a cutoff distance, and, since it is a parameter-free method, it is suitable for systems with inhomogeneous densities.

\subsection{Unsupervised learning}
\label{sec:ML}

\subsubsection{Nonlinear dimensionality reduction using neural-network-based autoencoders}
\label{sec:autoencoder}

\begin{figure}[]
\centering
\includegraphics[width=1.\linewidth]{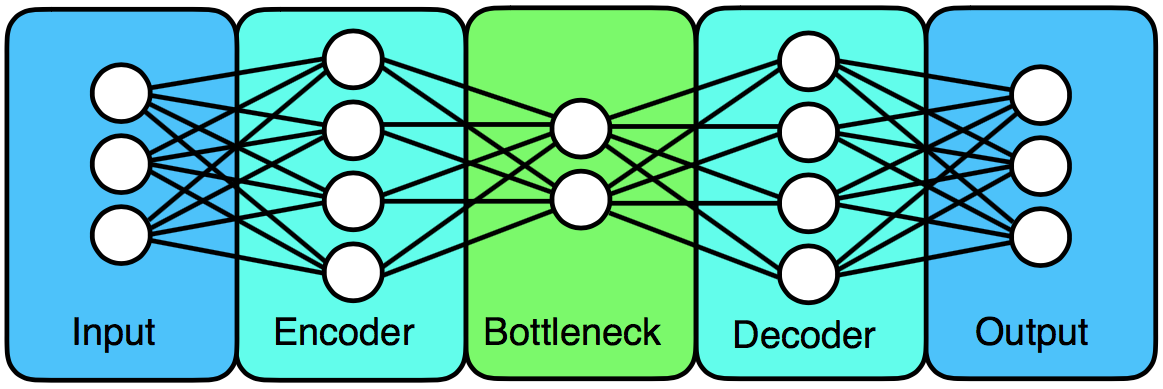}
\caption{Architecture of a neural-network based autoencoder. The encoder network finds a low-dimensional representation of the input, from which the decoder reconstructs an approximation of the input as output.} 
\label{NNauto}
\end{figure}

%%%% NEW VERSION %%%%
In order to extract the relevant information contained in the vectors $\mathbf{Q}(i)$, we use neural-network-based autoencoders\cite{Rumelhart1986, Kramer1991, Scholz2002, Goodfellow, Bishop}. An autoencoder is a neural network that is trained to perform the identity mapping, where the network inputs are reproduced at the output layer. The network may be viewed as consisting of two parts: an encoder network, which performs a nonlinear projection of the input data into a low-dimensional subspace, and a decoder network that attempts to reconstruct the input data from the low dimensional projection. This architecture is represented in Fig. \ref{NNauto}. 

By training the autoencoder to perform the input reconstruction task over an ensemble of training examples, the encoder is forced to learn a low-dimensional nonlinear projection that preserves the most relevant features of the data and from which the higher-dimensional inputs can be approximately reconstructed by the decoder. In this work, the training data are the vectors $\mathbf{Q}(i)$ (Eqs. \ref{input1} or \ref{input2}) evaluated from snapshots of colloidal systems obtained via computer simulations, and the autoencoder is trained to find a low-dimensional projection of such vectors by eliminating irrelevant and redundant information.

In the present context, we employ feedforward and fully-connected autoencoders like the one presented in Fig. \ref{NNauto}. %The number of input and output nodes, $d$, is specified by the dimensionality of the input data. 
%The input vectors, $\mathbf{Q}(i)\in\mathbb{R}^d$, are approximately reconstructed by the network in the output layer, $\hat{\mathbf{Q}}(i)\in\mathbb{R}^d$. 
The number of input and output nodes, $d$, is specified by the dimensionality of the input vectors, $\mathbf{Q}(i)\in\mathbb{R}^d$, which are approximately reconstructed by the network in the output layer, $\hat{\mathbf{Q}}(i)\in\mathbb{R}^d$. The bottleneck layer contains the low-dimensional projection to be learned by the encoder, $\mathbf{Y}(i)\in\mathbb{R}^c$, whose dimensionality is controlled by the number of bottleneck nodes, $c<d$. Nonlinearity is achieved by providing both the encoder and the decoder with a fully-connected hidden layer with a nonlinear activation function. Here, we set the number of nodes in the hidden layers to $10 d$ and use a hyperbolic tangent as the activation function. For the bottleneck and output layers, instead, a linear activation function is used.

The internal parameters of the autoencoder, i.e. weights $\mathbf{W}\equiv\{w_j\}$ and biases $\mathbf{B}\equiv\{b_k\}$, are optimized during the training by minimizing the reconstruction error of the input data over a training set of $N$ training examples. Specifically, we consider the mean squared error (MSE) with the addition of a weight decay regularization term\cite{Bishop} to control the magnitude of the network weights
\begin{equation}
\label{loss}
E(\mathbf{W}, \mathbf{B}; \{\mathbf{Q}(i)\})=\frac{1}{N}\sum_{i=1}^{N}\left\lVert\mathbf{Q}(i)-\hat{\mathbf{Q}}(i)\right\lVert^2 + \lambda\sum_{j=1}^{M}w^2_j,
\end{equation} 
where $M$ is the total number of weights, whose value depends on the dimension of the network, and we set $\lambda=10^{-5}$. The function in Eq. \ref{loss} is minimized using mini-batch stochastic gradient descent with momentum\cite{BackProp, Momentum, Bishop}.

The optimal number of nodes in the bottleneck layer, $c$, which defines the unknown relevant dimensionality of the input data, can be determined by computing the reconstruction MSE and looking for the existence of an elbow in the MSE as a function of $c$ \cite{chen2018collective}. For convenience, we rescale the MSE by the mean squared deviation (MSD) of the vectors $\mathbf{Q}(i)$,
\begin{equation}
\text{MSD}= \frac{1}{N}\sum_{i=1}^{N}\left\lVert\mathbf{Q}(i)-\bar{\mathbf{Q}}\right\lVert^2,
\end{equation}
where $\bar{\mathbf{Q}}$ is the mean input vector. To detect the presence of an elbow we use the L-method proposed by Salvador and Chan\cite{Lmethod}. For all systems examined in this work, we found a dimensionality of $c=2$ to be sufficient.

Once the autoencoder is trained, the encoder network alone is retained in order to perform the nonlinear mapping of the input vectors $\mathbf{Q}(i)$ onto the low-dimensional  subspace defined by the bottleneck layer, $\mathbf{Y}(i)$.

\subsubsection{Learning from the autoencoder}
\label{sec:learning}

One of the main advantages of using a neural-network-based autoencoder over other nonlinear techniques for dimensionality reduction is that it furnishes an exact analytical mapping (and an approximate inverse mapping) between the original input space and its low-dimensional projection. In finding such a mapping, the autoencoder must understand which of all the BOPs given as input in the vectors $\mathbf{Q}(i)$ are the most relevant for the system under analysis. Extrapolating this knowledge would help us understand the relevant symmetries distinguishing the different environments possibly present in the system.

Several methods to assess the relative importance of input variables in neural network models have been proposed\cite{Yao1998, Scardi1999, Gevrey2003, Olden2004}. Here, we consider the input perturbation\cite{Yao1998, Scardi1999, Gevrey2003, Olden2004} and the improved stepwise\cite{Gevrey2003, Olden2004} methods. Both techniques require the use of a single trained model, avoiding having to repeat the training of the autoencoder multiple times.

The input perturbation method assesses the variation in the MSE of the autoencoder by adding, in turn, a small amount of white noise to the $k$-th input, while holding all the other inputs at their observed values. Here, we set the white noise to $10\%$ and $50\%$ of each input, as suggested in Ref.\onlinecite{Gevrey2003}. The input variables whose changes affect the output the most, leading to a large increase in the MSE, are the ones that have the most relative influence.

The improved stepwise method is very similar in spirit, but instead of adding noise to one of the inputs, it replaces all its values with its mean over the whole dataset. Also in this case, the most relevant input variables are identified as the ones whose replacement causes the largest increase in the MSE.  

In both cases, a quantitative measure of the relative importance, $\text{RI}_k$, of the $k$-th input can be obtained as
\begin{equation}
\label{RI}
\text{RI}_k = \frac{\Delta E_k}{\sum_{j=1}^{d} \Delta E_j},
\end{equation}
where $\Delta E_k$ is the variation in the MSE caused by the change applied to the $k$-th input, and the sum in the denominator runs over all the input variables.

\subsubsection{Clustering}
\label{sec:clustering}
In order to cluster together similar environments in the low-dimensional subspace found by the encoder, we use Gaussian mixture models (GMMs) as implemented in scikit-learn\cite{scikit}. %GMM is a probabilistic model which attempts to create a probability density function that agrees well with the distribution of observed data by using a finite number of Gaussian functions. 

GMM is a probabilistic model that assumes that the observed data are generated from a mixture of a finite number of Gaussian distributions with unknown parameters. Such parameters are optimized iteratively with the expectation-maximization (EM) algorithm\cite{EM} in order to create a probability density function that agrees well with the distribution of the data. The number of Gaussian components in the mixture, $N_G$, is usually found by minimizing the Bayesian information criterion (BIC)\cite{BIC}, which measures how well a GMM fits the observed data while penalizing models with many parameters to prevent overfitting.  The output of a trained GMM is  a list of probabilities, $p_{ij}$, corresponding to the posterior probabilities of the $i$-th observation to arise from the $j$-th component in the mixture model. 

The simplest form of clustering that can be applied consists in considering each mixture component as generating a separate cluster and assigning each observation to the component with the highest posterior probability. However, while this procedure works perfectly for clusters that are really generated from a mixture of separate multivariate normal distributions, the clusters that underline our data are very often far from being Gaussian-distributed in space. As a consequence, a single cluster in the data may be detected as two or more mixture components (if its distribution is indeed better approximated by a mixture of Gaussians than by a single Gaussian function), meaning that the number of clusters in the data may in general be different from the number of components found by minimizing the BIC. 

To overcome this problem, we use the method proposed by Baudry \emph{et al.}\cite{Baudry2010}. The idea is to first use the BIC in order to find a GMM with $N_G$ components that fits the data well. Then, a sequence of candidate clusterings with $K=N_G, N_G-1,\dots,1$ clusters is formed by successively merging a pair of components. At each step, the two mixture components to be merged are chosen so as to minimize the entropy of the resulting clustering,  defined as 
\begin{equation}
S_K = -\sum_{i=1}^{N}\sum_{j=1}^{K}p_{ij}\ln(p_{ij}),
\end{equation}
where $N$ is the number of observations and $K$ the number of clusters. Finally, the optimal number of clusters is found by looking for the existence of an elbow in the entropy $S_K$ as a function of $K$. Again, we detect the elbow with the L-method of Salvador and Chan\cite{Lmethod}. 

This procedure autonomously finds the number of clusters underlying the data, corresponding to the distinct particle environments present in the system under analysis. Moreover, note that this is a soft clustering technique, meaning that each particle is not simply assigned to a cluster corresponding to a specific local environment,  but rather it has a certain probability of belonging to any of the identified clusters. As a result, particles whose environment is not well defined, such as can occur at interfaces, will have similar probabilities of belonging to different clusters. In the following, we will refer to these probabilities as membership probabilities.

\section{Results and Discussion}
In this section, we show how our method performs on snapshots of a wide range of colloidal systems obtained via computer simulations. We first present in detail the whole procedure for the analysis of a ``test'' example for which we compare the results with a more standard, system-specific, methodology. A shorter, more concise, discussion is dedicated to the results obtained for the other systems analyzed, including systems with grain boundaries, anisotropic hard cubes, and binary mixtures.

\subsection{Single-component hard spheres}
As a first test case we examine a snapshot from a Monte Carlo (MC) simulation of single-component hard spheres of diameter $\sigma$, which is shown in Fig. \ref{fig:snap1}. The simulation was performed in the canonical ensemble (constant number of particles $N$, volume $V$ and temperature $T$) and exhibited a coexistence between the fluid, hexagonal close-packed (HCP) and face-centered cubic (FCC) phases. The system contained $N=1536$ particles and was at a number density $\rho\sigma^3=1.01$. 

\subsubsection{Analysis}

\begin{figure*}
\centering
\subfloat[Snapshot under analysis.]{%
\resizebox*{0.45\linewidth}{!}{\includegraphics{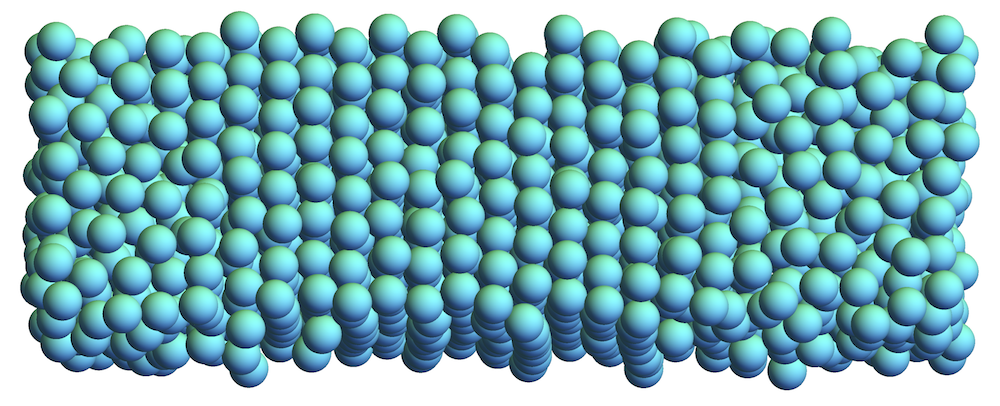}}\label{fig:snap1}}\hspace{0.5cm}\\ 
\subfloat[Rescaled MSE of the autoencoder as a function of the number of bottleneck nodes. Solid lines show the presence of an elbow at $c=2$.]{%
\resizebox*{0.45\linewidth}{!}{\includegraphics{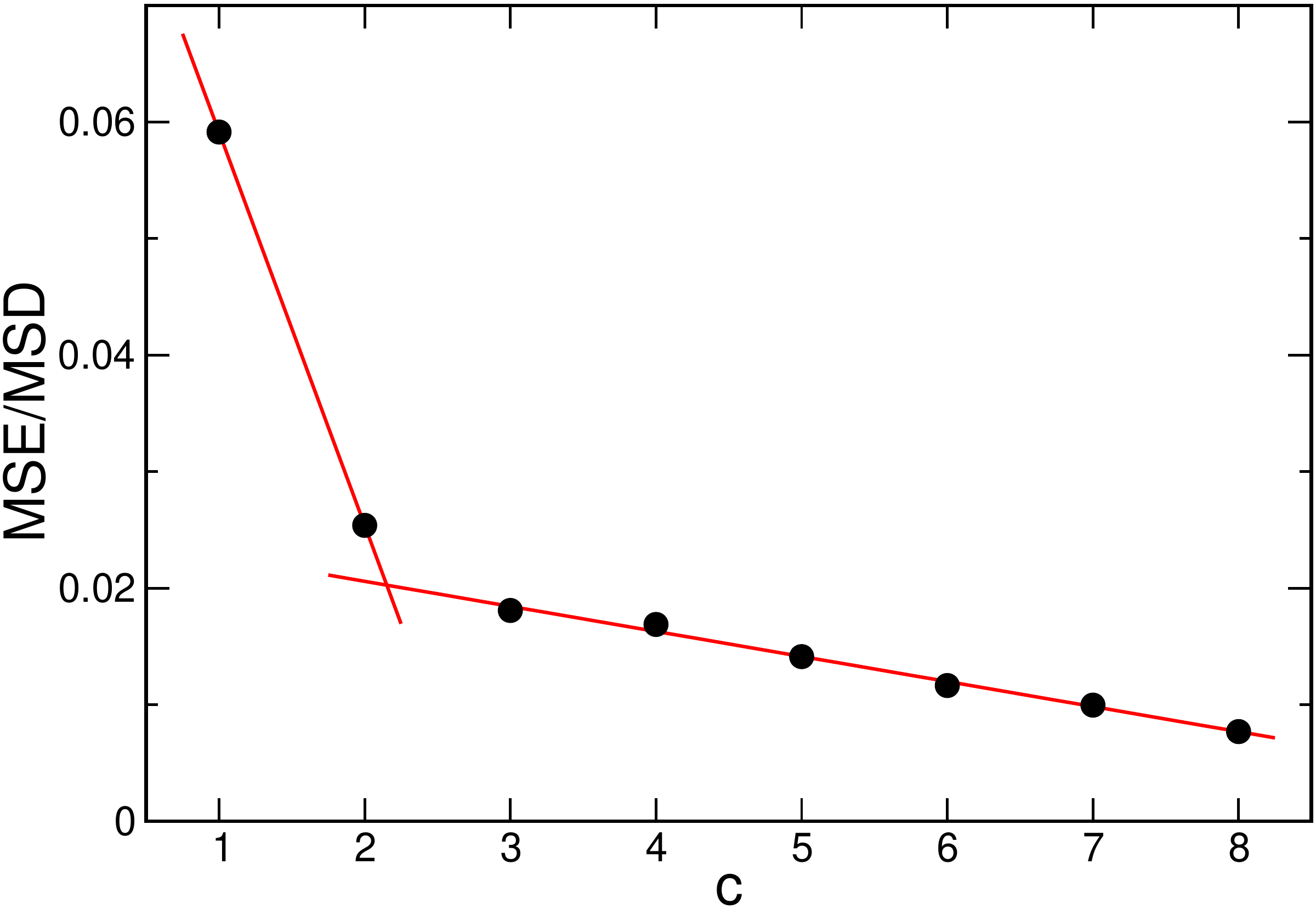}}\label{fig:dim}}\hspace{0.5cm}
\subfloat[BIC as a function of the number of components in the GMM. The minimum is highlighted in red. Dashed lines are only a guide for the eyes.]{%
\resizebox*{0.45\linewidth}{!}{\includegraphics{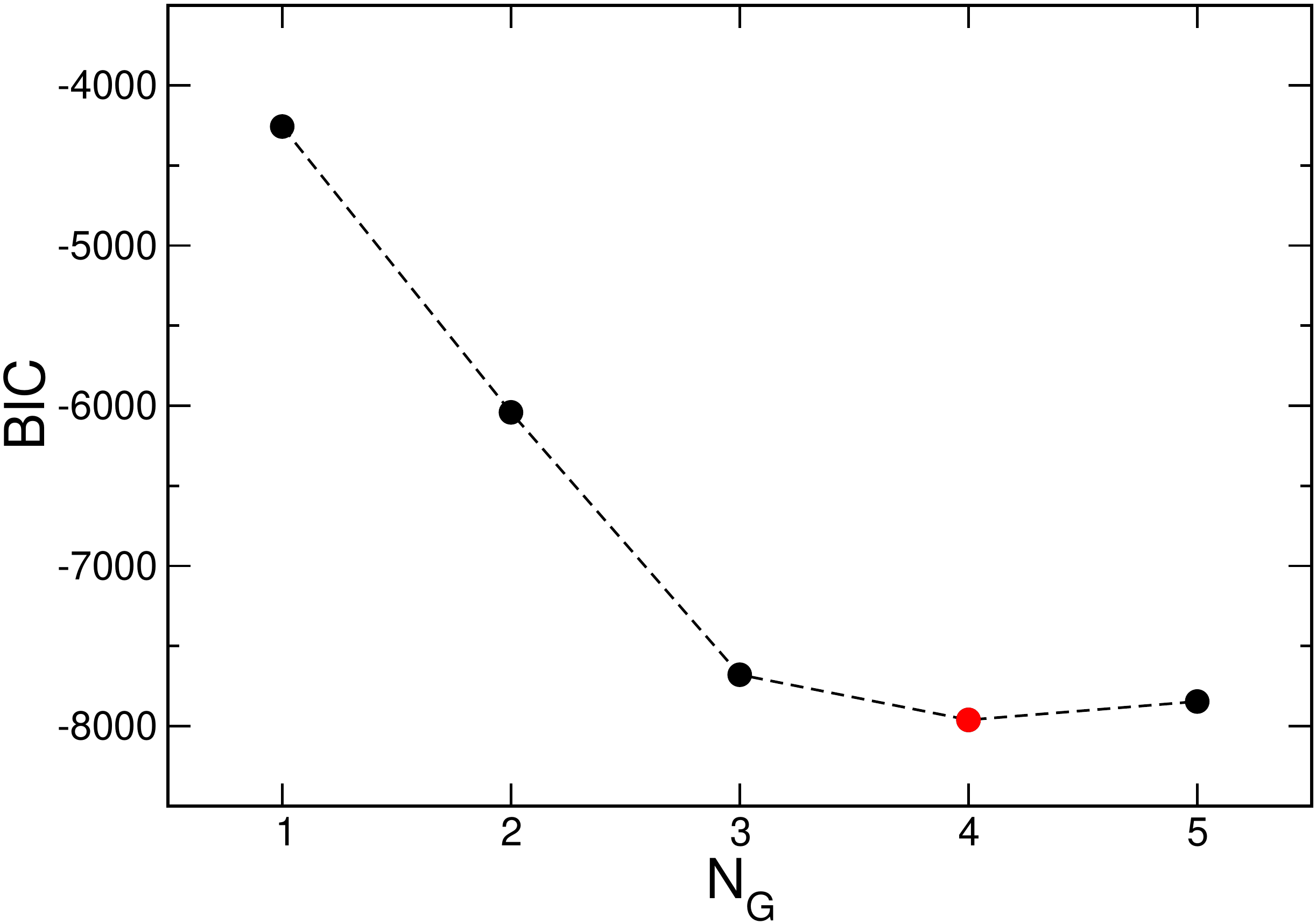}}\label{fig:bic}}\\
\subfloat[Entropy of the clustering as a function of the number of clusters. Solid lines show the presence of an elbow at $K=3$.]{%
\resizebox*{0.45\linewidth}{!}{\includegraphics{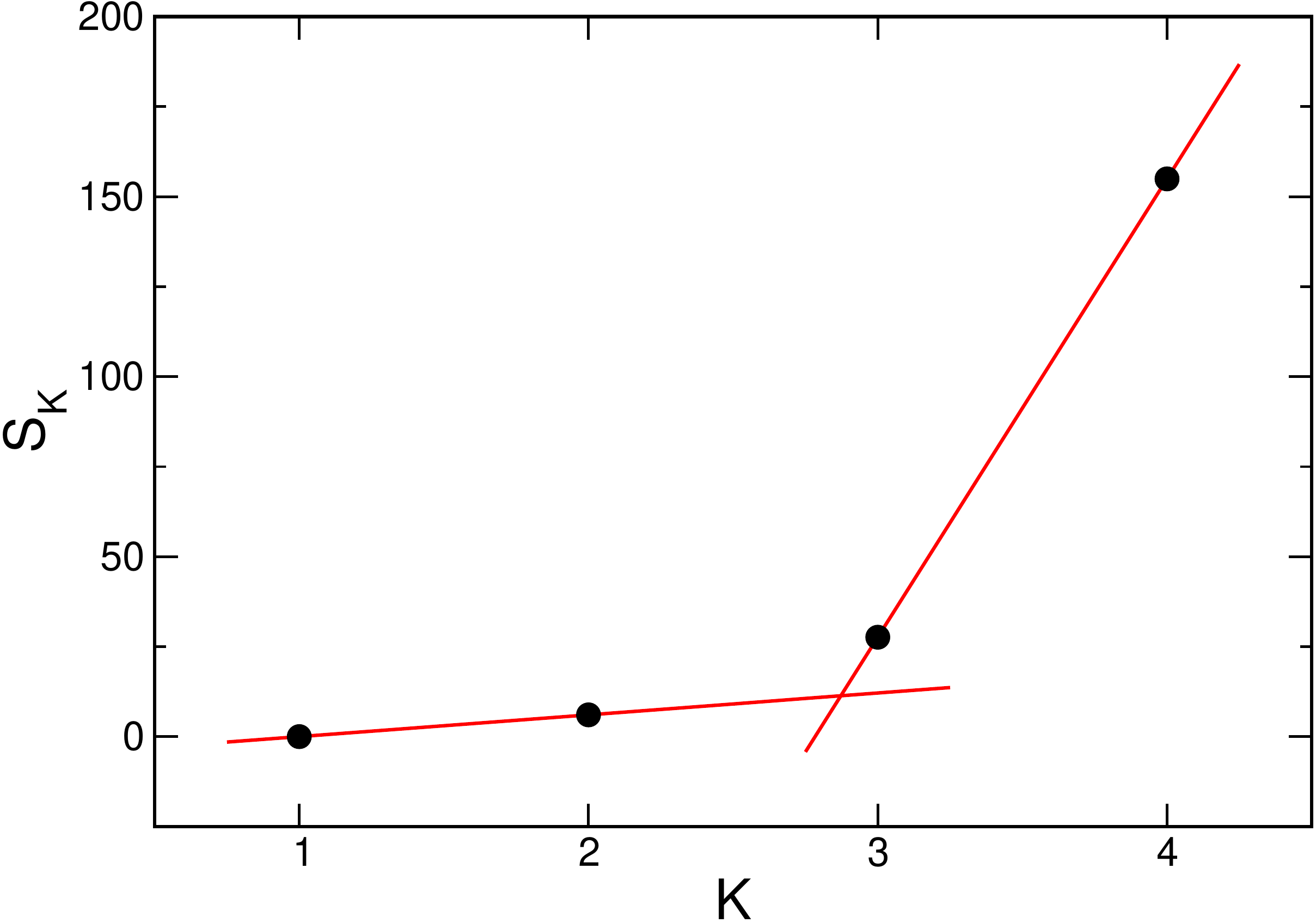}}\label{fig:entropy}}\hspace{0.5cm}
\subfloat[Projection of the vectors $\mathbf{Q}(i)$ onto the 2-dimensional space found by the encoder. Colors represent the distinct environments identified by the clustering.]{%
\resizebox*{0.45\linewidth}{!}{\includegraphics{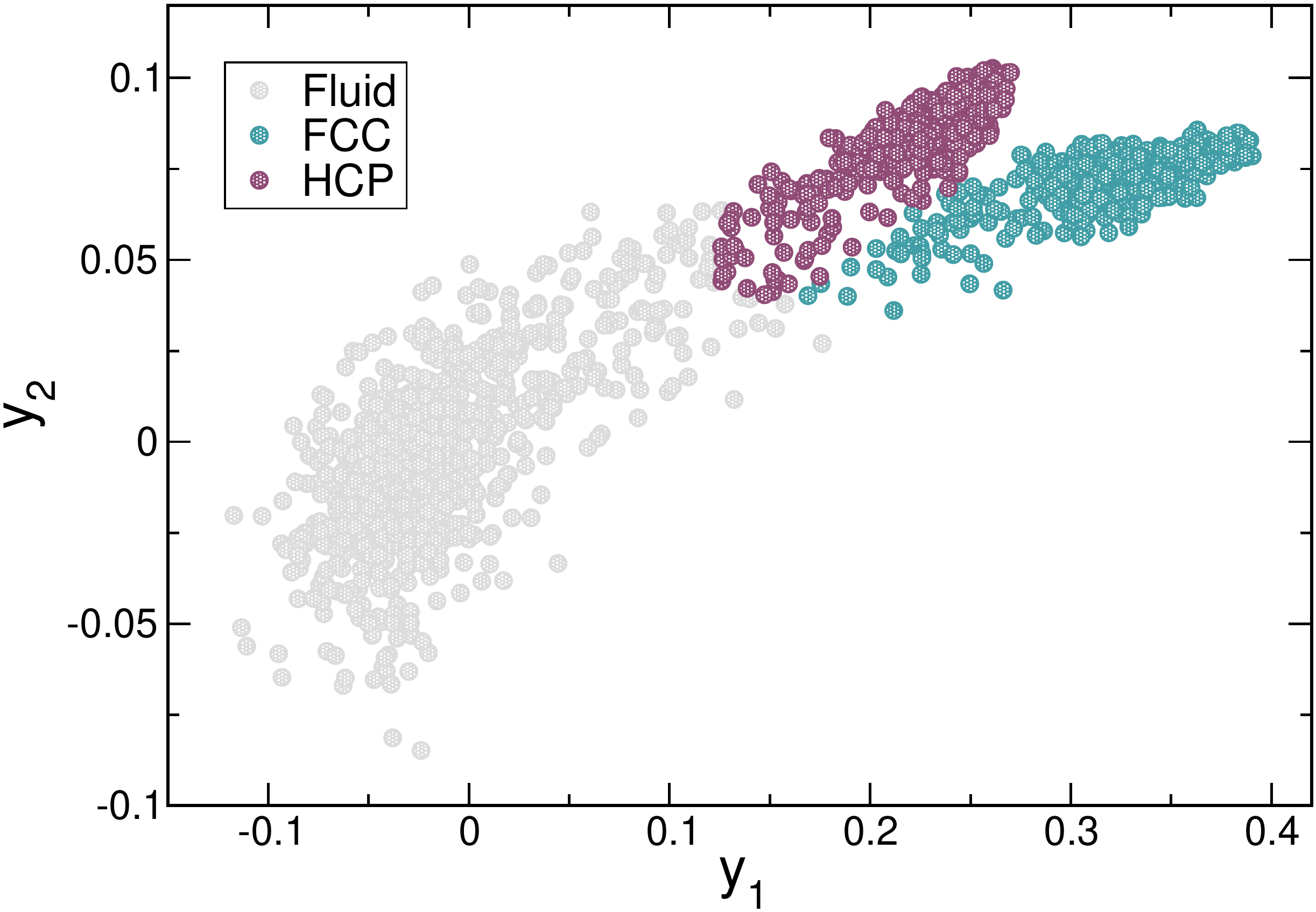}}\label{fig:scatter1}}\\
\subfloat[Classification of the snapshot in panel (a). The RGB color of each particle is a linear combination of the colors of the three phases (see legend) with the associated membership probabilities as coefficients.]{%
\resizebox*{0.49\linewidth}{!}{\includegraphics{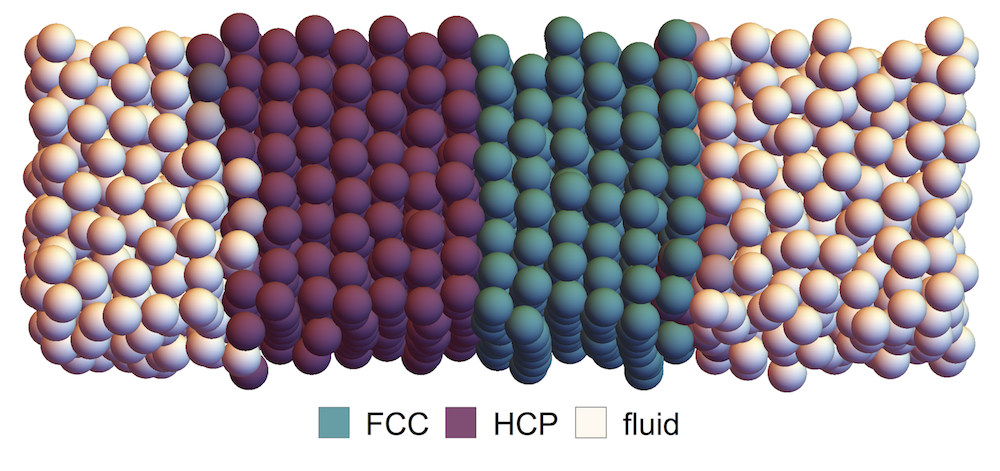}}\label{fig:snap1_un}}
\caption{Analysis of a snapshot from a MC simulation of hard spheres showing a coexistence between fluid, FCC and HCP crystals.}\label{fig:analysis}
\end{figure*}

Starting from the raw coordinates of each particle $i$ in the system, we build the vectors $\mathbf{Q}(i)$ in Eq. \ref{input1} and use them as an input for the autoencoder. To find the optimal dimensionality of the bottleneck layer, $c$, we evaluate the reconstruction MSE of the autoencoder  for $c\in[1,8]$. A plot of the rescaled MSE as a function of $c$ is shown in Fig. \ref{fig:dim}. Solid lines, obtained with the L-method, clearly show the presence of an elbow at $c=2$, indicating that a two-dimensional projection of the original input vectors is sufficient to preserve the relevant information. The projection learned by the encoder is depicted in Fig. \ref{fig:scatter1}, where each point corresponds to a particle in the system. Note that the colors in Fig. \ref{fig:scatter1} do not matter yet.

We then apply GMM in this two-dimensional space in order to identify the relevant clusters, i.e. the distinct particle environments. Following the method of Baudry \emph{et al.}, we first optimize the number of Gaussian components in the mixture model, $N_G$, by minimizing the BIC. 
The BIC as a function of $N_G$ is shown in Fig. \ref{fig:bic} and has a minimum for $N_G=4$. %Figure \ref{fig:bic} shows that the BIC has a minimum for $N_G=4$. 
Then, the optimal number of clusters, $K$, is found by successively merging a pair of components and looking for the existence of an elbow in the clustering entropy as a function of $K$ (see Fig. \ref{fig:entropy}). The elbow is detected at $K=3$, meaning that the unsupervised learning identifies three relevant environments. Note  that we know beforehand the three phases present in the system (FCC, HCP and fluid), so that we can easily associate each cluster with the correct phase. An idea of the partitioning of space performed by the clustering is given in Fig. \ref{fig:scatter1}, where the color of each point is determined by the cluster with the highest membership probability. As discussed in section \ref{sec:clustering}, however, some points might have a non-vanishing probability of belonging to more than one cluster. In order to fully account for such additional information, in the snapshot in Fig. \ref{fig:snap1_un} the RGB color of the particles is  obtained as a linear combination of the colors of the three clusters, with the associated membership probabilities as coefficients. As a result, particles with an environment falling at the boundary between two clusters, hence having two non-vanishing membership probabilities of similar magnitude, appear with a different color from those in the legend. This behavior is clearly observed for some of the particles at the crystal-fluid interfaces.
 
In the following, we compare the results obtained with a more standard methodology specifically tuned for distinguishing the phases in this system.

\subsubsection{Comparison}
\label{sec:comparison}
A common method to classify different phases on a single-particle basis consists in finding one (or more) pair(s) of BOPs, $(\bar{q}_l, \bar{q}_{l'})$, whose distributions in the phases of interest are considerably different, so that it is possible to define separate regions in the $\bar{q}_l$-$\bar{q}_{l'}$-plane corresponding to different particle environments. %Then, particles are classified depending on the region in which their BOPs fall. 
The pair of BOPs to consider strongly depends on the environments one wishes to distinguish, and it is usually found by trial and error.

\begin{figure}[h!]
\centering
\subfloat[]{%
\resizebox*{0.8\linewidth}{!}{\includegraphics{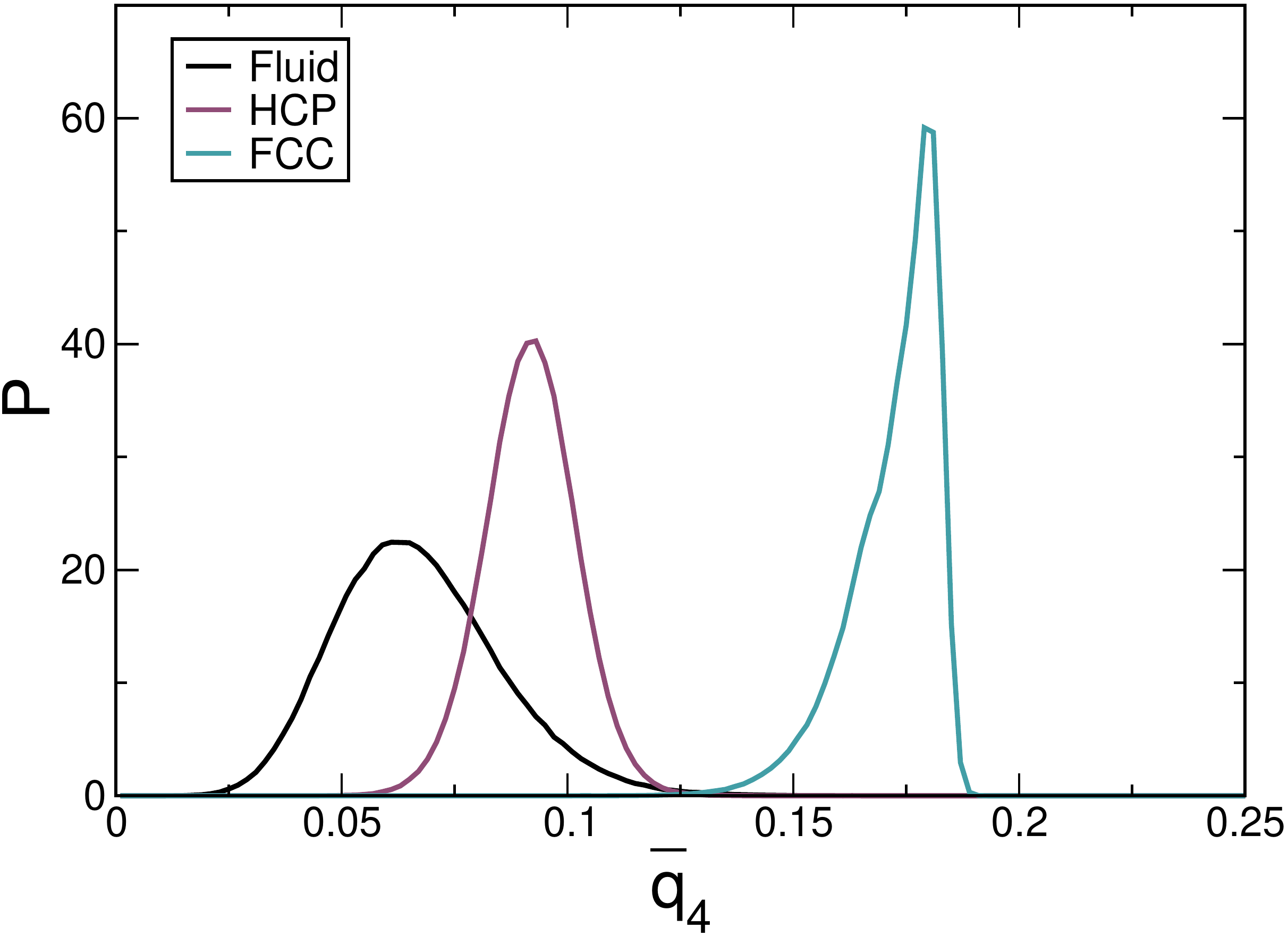}}\label{fig:q4}}\hspace{0.5cm}
\subfloat[]{%
\resizebox*{0.8\linewidth}{!}{\includegraphics{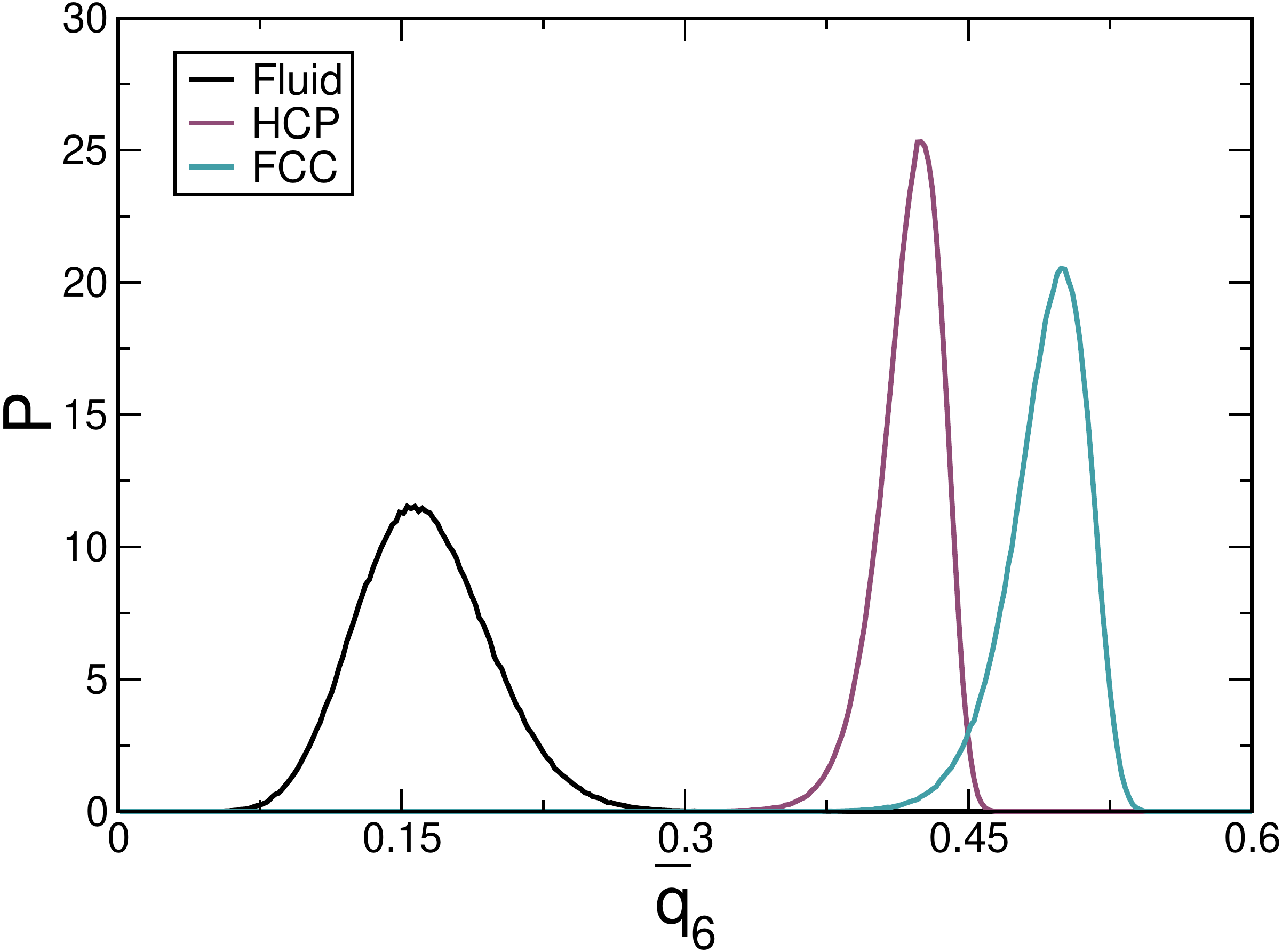}}\label{fig:q6}}\hspace{0.5cm}
\caption{(a) Probability distribution of $\bar{q}_4$ for the fluid, FCC and HCP hard-sphere phases. (b) Probability distribution of $\bar{q}_6$ for the same phases.}\label{fig:distributions}
\end{figure}

\begin{figure}[h!]
\centering
\includegraphics[width=1.\linewidth]{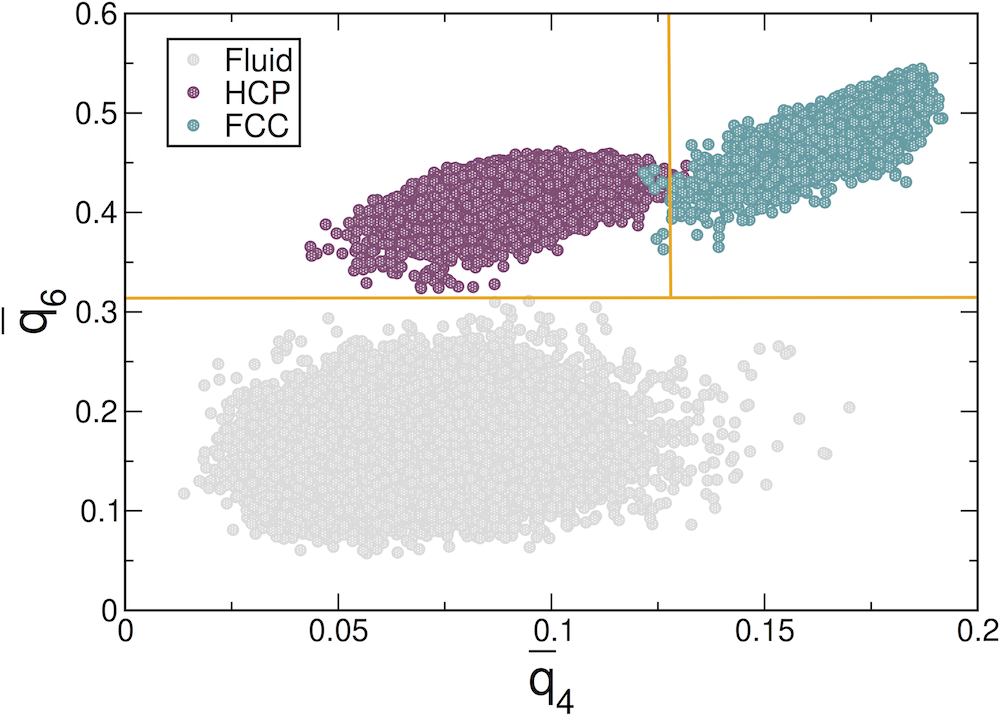}
\caption{Comparison between the $\bar{q}_4$-$\bar{q}_6$-plane for the fluid, FCC and HCP hard-sphere phases. Each point correspond to a particular particle. 20000 points from each phase were chosen randomly. Orange solid lines show a possible linear separation of the three phases in this plane.}
\label{fig:q4q6}
\end{figure}

For the FCC, HCP and fluid phases, $\bar{q}_4$ and $\bar{q}_6$ are the most common choice\cite{Lechner2008}. Fig. \ref{fig:distributions} shows the probability distribution of $\bar{q}_4$ (\ref{fig:q4}) and $\bar{q}_6$ (\ref{fig:q6}) for the three hard-sphere phases, obtained from separate MC simulations in the canonical ensemble. Simulations of the two crystal phases were performed at a number density just above the coexistence region, $\rho\sigma^3=1.05$, while for the fluid phase we used a number density of $\rho\sigma^3=0.86$. 

\begin{figure}[]
\centering
\includegraphics[width=1.\linewidth]{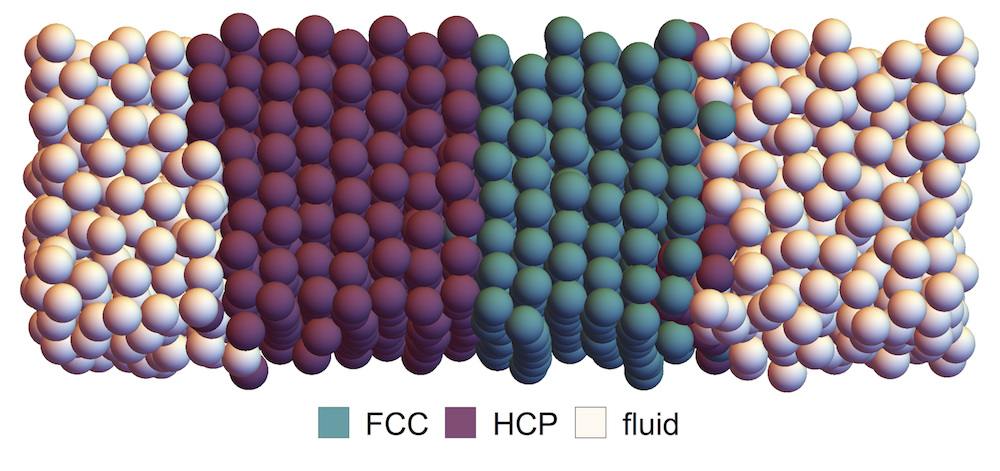}
\caption{Classification of the snapshot in Fig. \ref{fig:analysis} based on $\bar{q}_4$ and $\bar{q}_6$.} 
\label{fig:snap1_q4q6}
\end{figure}

The distributions of $\bar{q}_4$ in the two crystals are well separated, while the $\bar{q}_4$ distribution of the fluid phase strongly overlaps the one of the HCP crystal. On the other hand, the $\bar{q}_6$ distributions show a large separation between the fluid and the crystal phases, but a small overlap between the two crystals. Alone, none of the BOPs considered completely separates the three phases. However, a separation can be found in the $\bar{q}_4$-$\bar{q}_6$-plane. A comparison of the phases in the $\bar{q}_4$-$\bar{q}_6$-plane is shown in Fig. \ref{fig:q4q6}. In this plane, one can easily identify three linearly-separated regions associated with the three phases. A possible choice of such a separation is represented by the solid lines in Fig. \ref{fig:q4q6}. Based on this definition, the particles in Fig. \ref{fig:snap1} can be classified according to the region in which their BOPs fall. The results of this classification are presented in Fig. \ref{fig:snap1_q4q6} and they are in excellent agreement with the ones obtained via unsupervised learning (see Fig. \ref{fig:snap1_un} for comparison).  As can be seen by comparing Figs \ref{fig:snap1_un} and \ref{fig:snap1_q4q6}, the only differences between the two classifications are at the interfaces, and are generally particles for which the unsupervised learning algorithm gave at least two comparable membership probabilities (e.g. identified large probabilities of being in both fluid and FCC).

Note that the method presented above requires, in general, prior knowledge of the phases to be expected in the system under analysis. Moreover, additional simulations of such phases must be performed in order to (i) identify the relevant BOPs and (ii) define separations between the distinct particle environments. Both tasks, (i) and (ii), are autonomously performed by our unsupervised-learning method  based \emph{only} on the vectors of BOPs evaluated from the snapshot under study. 

Regarding the identification of the relevant BOPs, i.e. task (i), in the following section we present how such information can be extracted from the trained autoencoder.

\subsubsection{Learning from the autoencoder}
\begin{figure}[]
\centering
\includegraphics[width=1.\linewidth]{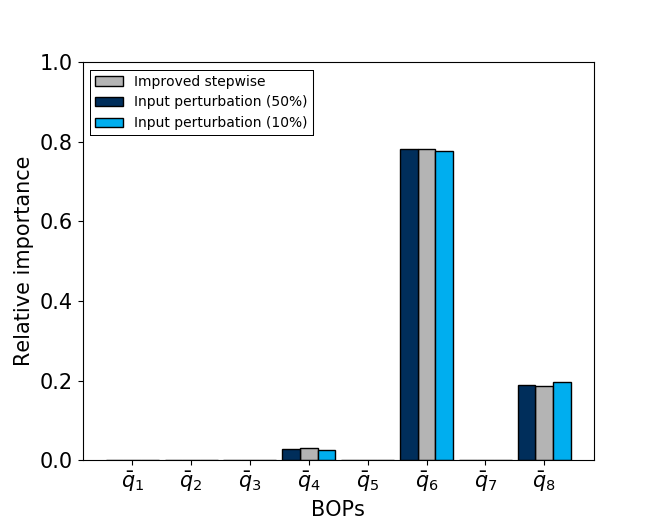}
\caption{Relative importance of the BOPs for the system in Fig. \ref{fig:analysis} assessed with the input perturbation and the improved stepwise methods. The input perturbation method is applied with two different amounts of white noise: $10\%$ and $50\%$ of the input, respectively.} 
\label{fig:RI}
\end{figure}

As discussed in section \ref{sec:learning}, several methods to assess the relative importance of input variables in neural network models are available. Here, we employ the input perturbation and the improved stepwise methods in order to understand which BOPs  were considered to be the most relevant by the autoencoder for the system in Fig. \ref{fig:analysis}. The relative importance of the BOPs evaluated with these methods is shown in Fig. \ref{fig:RI}. 

\begin{figure}[h]
\centering
\includegraphics[width=0.8\linewidth]{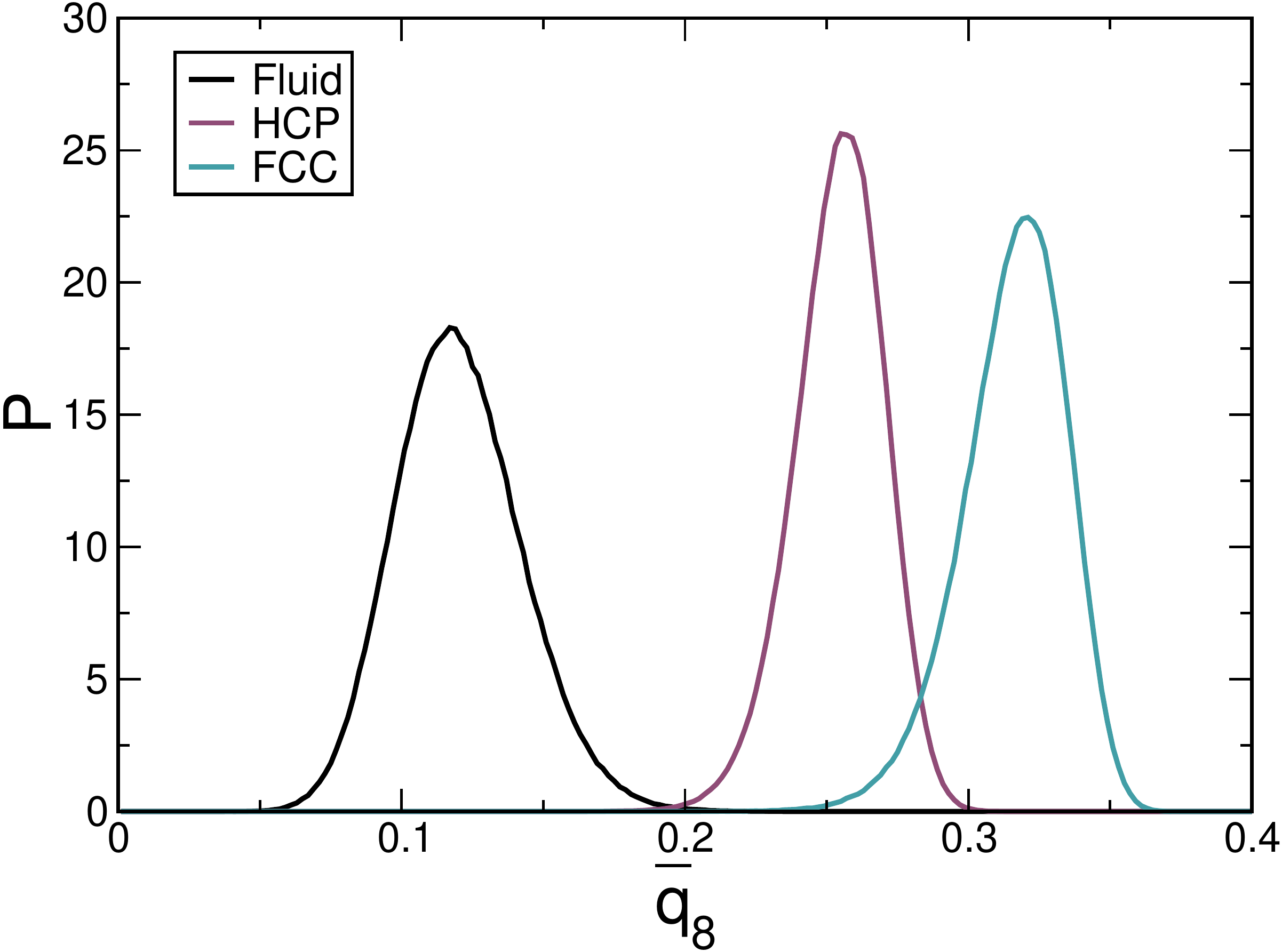}
\caption{Probability distribution of $\bar{q}_8$ for the fluid, FCC and HCP hard-sphere phases.}
\label{fig:q8}
\end{figure}

Only a small subset of three BOPs is found to be relevant and, as expected, $\bar{q}_6$ (RI$\sim 78\%$) and $\bar{q}_4$ (RI$\sim 3\%$) are part of it. Interestingly however, $\bar{q}_8$, which to our knowledge has never been used in literature, appears to be more important (RI$\sim 19\%$) than $\bar{q}_4$. To understand why this is, we evaluated the $\bar{q}_8$ distributions in the FCC, HCP and fluid hard-sphere phases from several snapshots (see Fig. \ref{fig:q8}). Such distributions are very similar to those obtained for $\bar{q}_6$, in the sense that they show a clear separation between the fluid and the crystal phases and only a small overlap between the two crystals. If we now go back to the snapshot shown in Fig. \ref{fig:snap1_un}, which is roughly half fluid and half crystal, then it is easy to understand why $\bar{q}_4$ has a lower relative importance than  $\bar{q}_6$ and $\bar{q}_8$.  

% Considering that the system under analysis is roughly half fluid and half crystal, it is easy to understand why $\bar{q}_4$ has a lower lower relative importance than  $\bar{q}_6$ and $\bar{q}_8$.

\subsection{Grain boundaries}
\begin{figure*}
\centering
\subfloat[]{%
\resizebox*{0.4\linewidth}{!}{\includegraphics{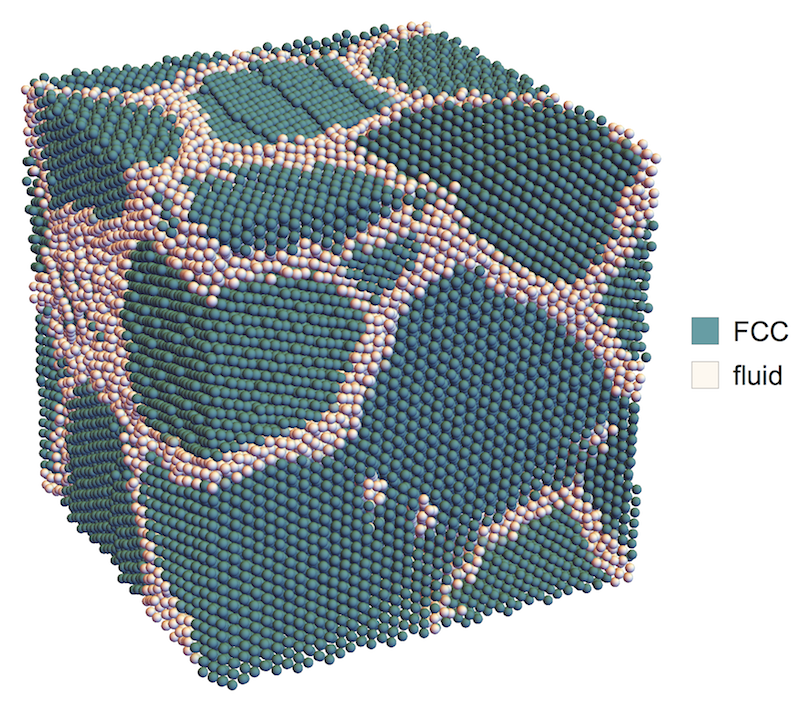}}\label{fig:snap_gb}}\hspace{1.0cm}
\subfloat[]{%
\resizebox*{0.45\linewidth}{!}{\includegraphics{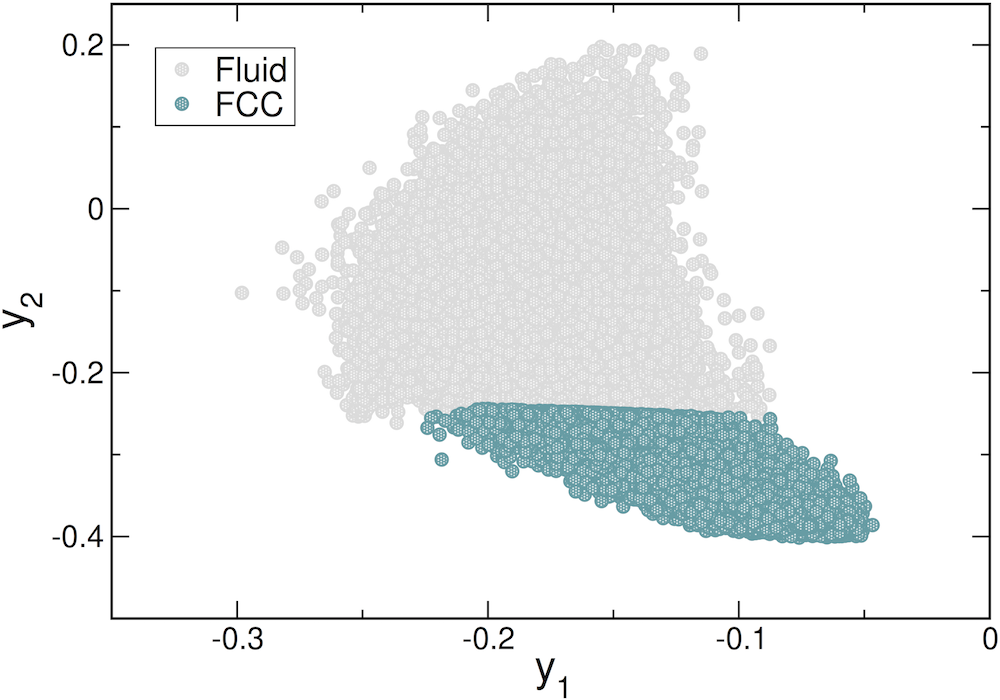}}\label{fig:scatter_gb}}\\
\caption{Analysis of a snapshot of a system with grain boundaries. (a) Classification of the snapshot under analysis. The RGB color of each particle is a linear combination of the colors of the two phases identified with the associated membership probabilities as coefficients. (b) Projection of the vectors $\mathbf{Q}(i)$ onto the 2-dimensional space found by the encoder. Colors represent the distinct environments identified by the clustering.}\label{fig:gb}
\end{figure*}

We now consider a snapshot of a system with FCC crystalline domains separated by grain boundaries, depicted in Fig. \ref{fig:snap_gb}. The system contains $N=83481$ particles interacting via the purely repulsive Weeks-Chandler-Andersen (WCA) potential
\begin{equation}
\label{eq:wca1}
\beta U_{WCA}(r)=
\begin{cases}
4\beta\epsilon\left[\left(\frac{\sigma}{r}\right)^{12}-\left(\frac{\sigma}{r}\right)^{6} +\frac{1}{4}\right], & \frac{r}{\sigma}\le 2^{1/6}\\
0, & \frac{r}{\sigma}>2^{1/6}
\end{cases}
\end{equation}
with $\sigma$ the particle diameter, $\beta\epsilon=40$ the energy scale, and $\beta=1/k_BT$, where $k_B$ is the Boltzmann constant and $T$ is the temperature. More details about the simulation can be found in Ref. \onlinecite{gb}.

The results of the unsupervised learning algorithm are summarized in Fig. \ref{fig:gb}.  Specifically,  Fig. \ref{fig:scatter_gb} shows the two-dimensional projection of the vectors $\mathbf{Q}(i)$ found by the encoder and the results of the clustering performed in this space.  Our method identifies two distinct particle environments, corresponding to particles within the grain boundaries and within the FCC crystalline domains. Note that the particles in the grain boundaries appear disordered and fluid-like.  This identification is used to colour the particles in Figure \ref{fig:snap_gb}. The most relevant BOPs in this system according to the autoencoder analysis are $\bar{q}_6$ (RI$\sim 78\%$) and $\bar{q}_8$ (RI$\sim 19\%$), which are indeed those whose distributions show the largest separation between the fluid and crystal phases.

% while the final classification is presented in Fig. \ref{fig:snap_gb}
% The unsupervised learning method identifies two distinct particle environments, corresponding to particles within the grain boundaries and within the FCC crystalline domains. Fig. \ref{fig:scatter_gb} shows the two-dimensional projection of the vectors $\mathbf{Q}(i)$ found by the encoder and the results of the clustering performed in this space, while the final classification is presented in Fig. \ref{fig:snap_gb}. The most relevant BOPs in this system according to the autoencoder analysis are $\bar{q}_6$ (RI$\sim 78\%$) and $\bar{q}_8$ (RI$\sim 19\%$), which are indeed those whose distributions show the largest separation between the fluid and crystal phases.
% 

\subsection{Hard cubes}
\begin{figure*}
\centering
\subfloat[]{%
\resizebox*{0.4\linewidth}{!}{\includegraphics{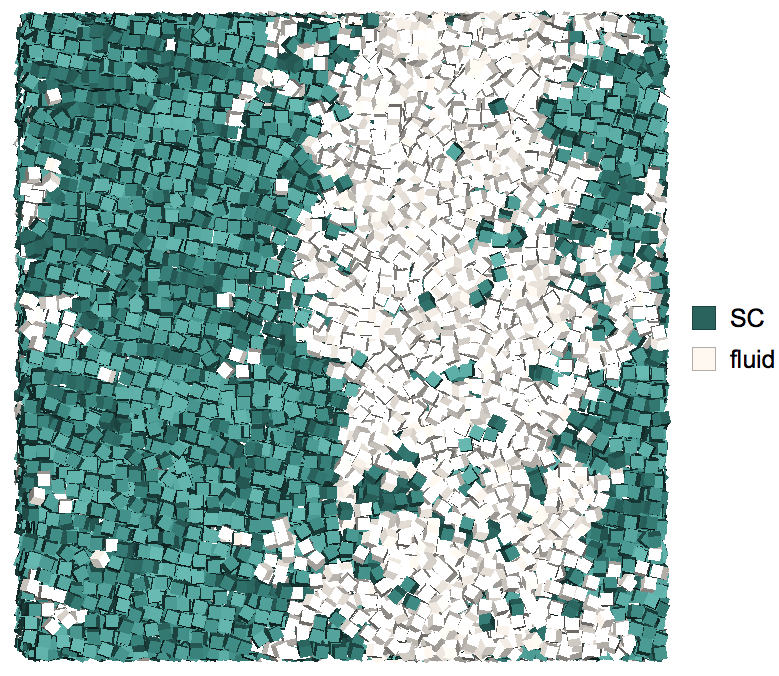}}\label{fig:snap_cubes}}\hspace{1.0cm}
\subfloat[]{%
\resizebox*{0.45\linewidth}{!}{\includegraphics{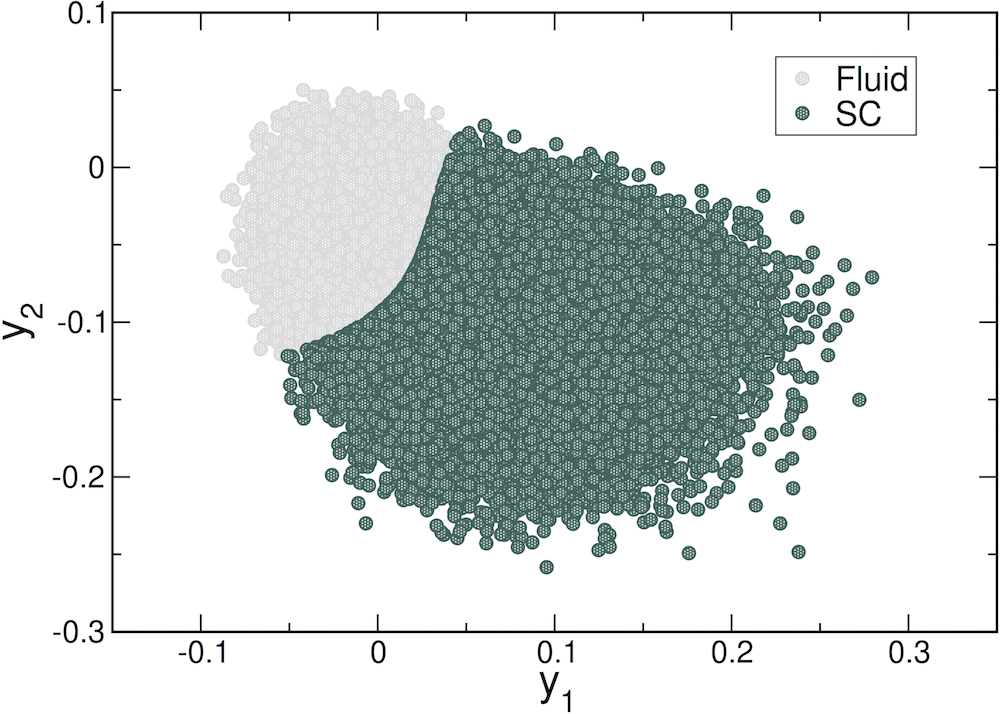}}\label{fig:scatter_cubes}}\\
\caption{Analysis of a snapshot of hard cubes showing a coexistence between the fluid and the SC crystal phases. (a) Classification of the snapshot under analysis. The RGB color of each particle is a linear combination of the colors of the two phases identified with the associated membership probabilities as coefficients. (b) Projection of the vectors $\mathbf{Q}(i)$ onto the 2-dimensional space found by the encoder. Colors represent the distinct environments identified by the clustering.}\label{fig:cubes}
\end{figure*}

The systems examined so far were all characterized by isotropic interactions between their constituents. As a further test of the performance and generality of our method, we now consider a snapshot of a system of hard cubes with edge length $\sigma$ (see Fig. \ref{fig:snap_cubes}), obtained from an event-driven molecular dynamics simulation (EDMD) in the canonical ensemble \cite{cubes}. The simulation was performed with $N=64000$ particles starting from a simple cubic (SC) crystal configuration at a number density in the fluid-crystal coexistence region, $\rho\sigma^3=0.475$. More details on the simulation can be found in Ref. \onlinecite{cubes}.

The two-dimensional projection of the vectors $\mathbf{Q}(i)$ found by the encoder and the results of the clustering performed in this space are shown in Fig. \ref{fig:scatter_cubes}, while the final classification is presented in Fig. \ref{fig:snap_cubes}. In agreement with the standard order parameters used in Ref. \onlinecite{cubes}, the unsupervised learning method identifies two distinct particle environments, corresponding to the fluid and SC crystal phases. From the autoencoder analysis we found that the most relevant BOPs for this system are $\bar{q}_6$ (RI$\sim 48\%$), $\bar{q}_4$ (RI$\sim 36\%$) and $\bar{q}_8$ (RI$\sim 16\%$).

\subsection{Binary mixture}
\begin{figure*}
\centering
\subfloat[]{%
\resizebox*{0.4\linewidth}{!}{\includegraphics{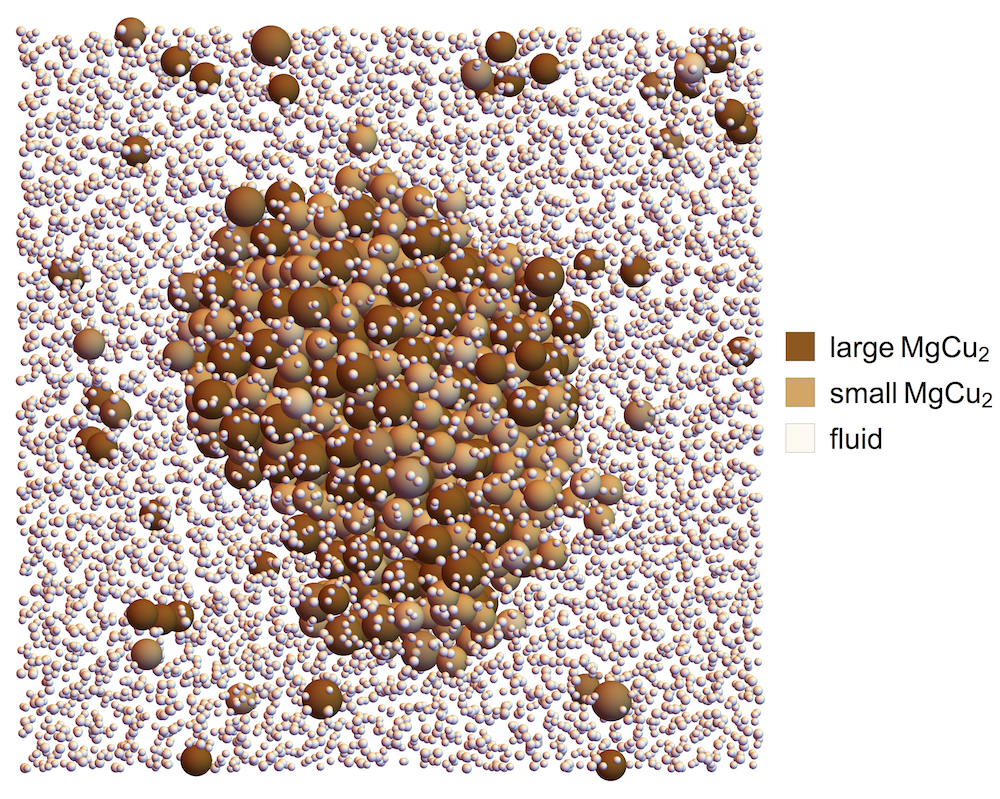}}\label{fig:snap_binnuc}}\hspace{1.0cm}
\subfloat[]{%
\resizebox*{0.45\linewidth}{!}{\includegraphics{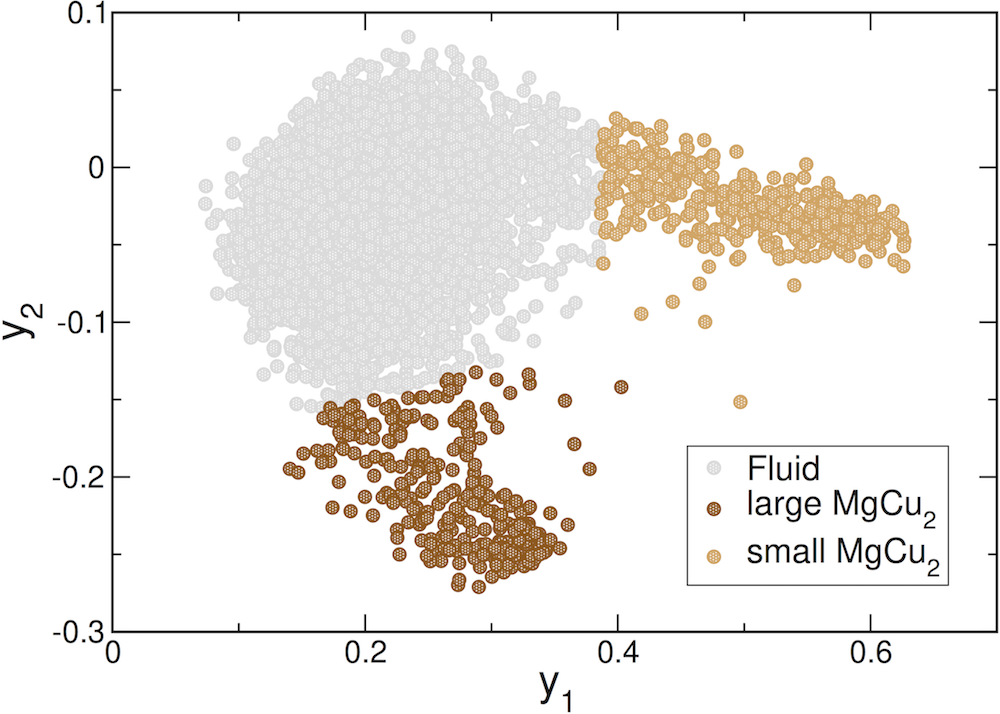}}\label{fig:scatter_binnuc}}\\
\caption{Analysis of a snapshot of a binary mixture showing a coexistence between the fluid phase and the MgCu$_2$ Laves phase. (a) Classification of the snapshot under analysis. The RGB color of each particle is a linear combination of the colors of the three local environments identified with the associated membership probabilities as coefficients. Fluid-like particles are displayed at $1/4$ of their actual size. (b) Projection of the vectors $\mathbf{Q}(i)$ onto the 2-dimensional space found by the encoder. Colors represent the distinct environments identified by the clustering.}\label{fig:binnuc}
\end{figure*}

After having considered single-component systems with both isotropic and anisotropic interactions, we now examine a binary mixture of large (L) and small (S) spheres of diameters $\sigma_L$ and $\sigma_S$, respectively, with a size ratio $\sigma_S/\sigma_L=0.78$ and a stoichiometry $N_s/N=\frac{2}{3}$. The particles in this system interact via the WCA potential, $U_{\alpha\gamma}(r)$, between species $\alpha=L,S$ and $\gamma=L,S$
\begin{equation}
\label{eq:wca2}
\beta U_{\alpha\gamma}(r)=
\begin{cases}
4\beta\epsilon\left[\left(\frac{\sigma_{\alpha\gamma}}{r}\right)^{12}-\left(\frac{\sigma_{\alpha\gamma}}{r}\right)^{6} +\frac{1}{4}\right], & \frac{r}{\sigma_{\alpha\gamma}}\le 2^{1/6}\\
0, & \frac{r}{\sigma_{\alpha\gamma}}>2^{1/6}
\end{cases}
\end{equation}
where $\sigma_{\alpha\gamma}=(\sigma_{\alpha}+\sigma_{\gamma})/2$ and $\beta\epsilon$ is the energy scale. The snapshot considered here, depicted in Fig. \ref{fig:snap_binnuc}, was obtained from a MC simulation  in the isothermal-isobaric ensemble (constant number of particles $N$, pressure $P$ and temperature $T$) with $N=8073$ particles, $\beta\epsilon = 5$ and $\beta P\sigma_L^3= 24.2$, and shows a coexistence between the MgCu$_2$ Laves phase and the fluid phase. More details about the simulation can be found in Ref. \onlinecite{dasgupta2019softness}.

Recall that, since we are dealing with a binary mixture, we describe the local environment of each particle $i$ in the system with a 16-dimensional vector of BOPs, $\mathbf{Q}(i) = (\{\bar{q}_l(i)\}, \{\bar{q}^{ss}_l(i)\})$, where the first set of 8 BOPs, $\{\bar{q}_l(i)\}$ with $l=1,\dots,8$, are evaluated considering all the nearest neighbors of particle $i$, while the second set of 8 BOPs, $\{\bar{q}^{ss}_l(i)\}$, is evaluated considering only the nearest neighbors of the same species ($s=L$ or $s=S$) as particle $i$.  

The results of the unsupervised learning classification are summarized in Fig. \ref{fig:binnuc}.  Specifically,  Fig. \ref{fig:scatter_binnuc} shows the two-dimensional projection of the vectors $\mathbf{Q}(i)$ found by the encoder and the results of the clustering performed in this space.  Our method identifies three distinct particle environments, corresponding to the large ($L$) and small ($S$) particles in the MgCu$_2$ Laves phase  and the particles in the fluid phase. Note that since large and small particles in the fluid phase appear equally disordered, the algorithm groups them together in the same cluster. This identification is used to colour the particles in Figure \ref{fig:snap_binnuc}.

\begin{figure}[]
\centering
\includegraphics[width=1.\linewidth]{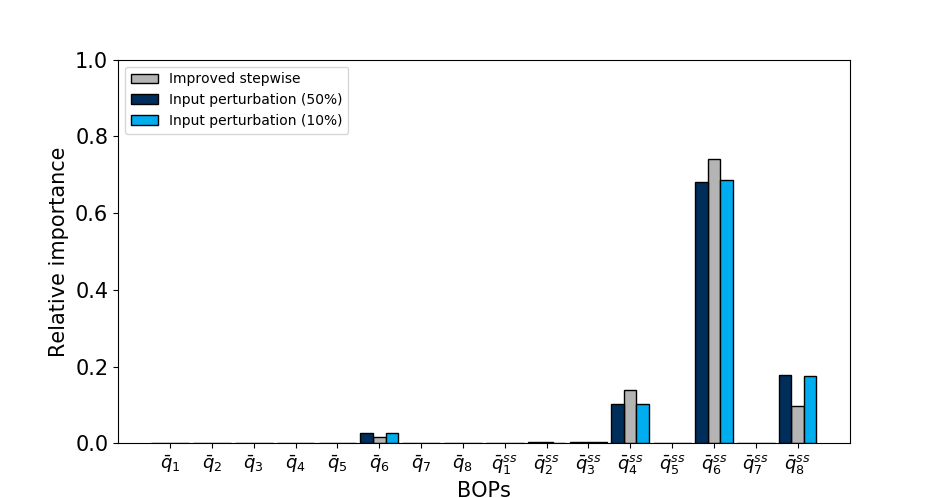}
\caption{Relative importance of the BOPs for the system in Fig. \ref{fig:binnuc} assessed with the input perturbation and the improved stepwise methods. The input perturbation method is applied with two different amounts of white noise: $10\%$ and $50\%$ of the input, respectively.} 
\label{fig:RI2}
\end{figure}

The relative importance of the BOPs obtained from the autoencoder analysis is shown in Fig. \ref{fig:RI2}. Interestingly, the BOPs evaluated considering all the nearest neighbors of the particles are not found to be relevant for this system, with the only exception of $\bar{q}_6$ (RI$\sim 2\%$). The largest variations in the environments are instead found in three of the BOPs evaluated considering only the nearest neighbors of the same species as the reference particle, specifically $\bar{q}^{ss}_6$ (RI$\sim 70\%$), $\bar{q}^{ss}_8$ (RI$\sim 15\%$), and  $\bar{q}^{ss}_4$ (RI$\sim 11\%$).

\subsection{Nucleation and crystal growth}
\begin{figure*}[]
\centering
\includegraphics[width=1.\linewidth]{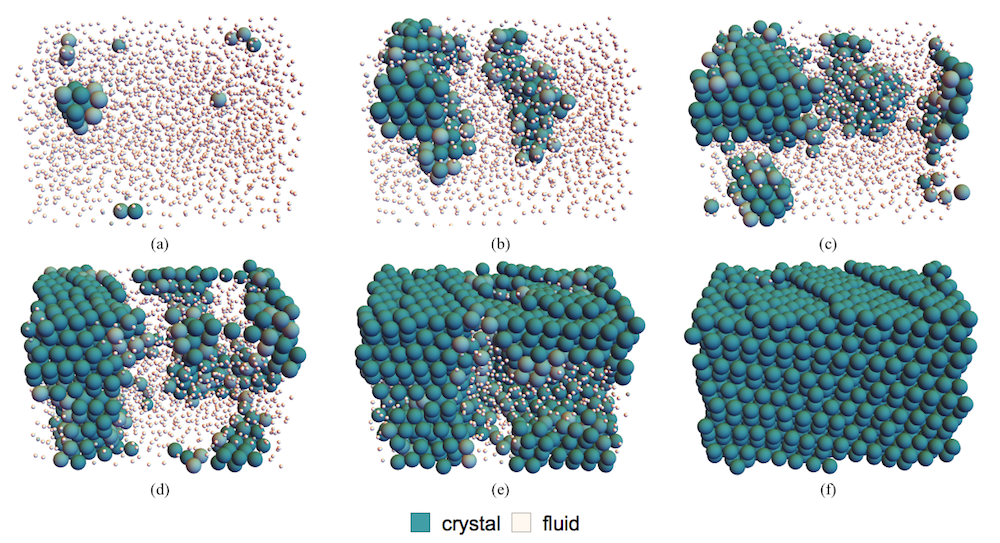}
\caption{Classification of snapshots of a MC simulation with $N=2048$ particles interacting via the WCA potential, showing a transition from the fluid to the crystal phase. The RGB color of each particle is a linear combination of the colors of the two phases identified with the associated membership probabilities as coefficients. Fluid-like particles are displayed at $1/4$ of their actual size.} 
\label{fig:snap_wca_nuc}
\end{figure*}

Finally, we examine the nucleation and crystal growth of a single-component system of particles with diameter $\sigma$, interacting via the WCA potential (Eq. \ref{eq:wca1}). Note that the nucleation and phase behavior of this system has been extensively studied in Ref. \onlinecite{wca}. Here, we performed a MC simulation in the isothermal-isobaric ensemble with $N=2048$ particles, $\beta\epsilon = 40$ and $\beta P\sigma^3= 30$. Six snapshots from this simulation, showing the transition from the fluid to the crystal phase, are depicted in Fig. \ref{fig:snap_wca_nuc}.

We started our structural analysis by considering all snapshots together.  For all particles, we evaluated the vectors $\mathbf{Q}(i)$ which we used as the input of the unsupervised learning algorithm. Performing a single analysis for all the snapshots in Fig. \ref{fig:snap_wca_nuc} guarantees that: (i) a sufficient statistics of the two environments, i.e. fluid and crystalline, is included in the analysis, and (ii) the same clustering is applied to each snapshot, allowing a quantitative comparison between them. 

\begin{figure}[]
\centering
\includegraphics[width=1.\linewidth]{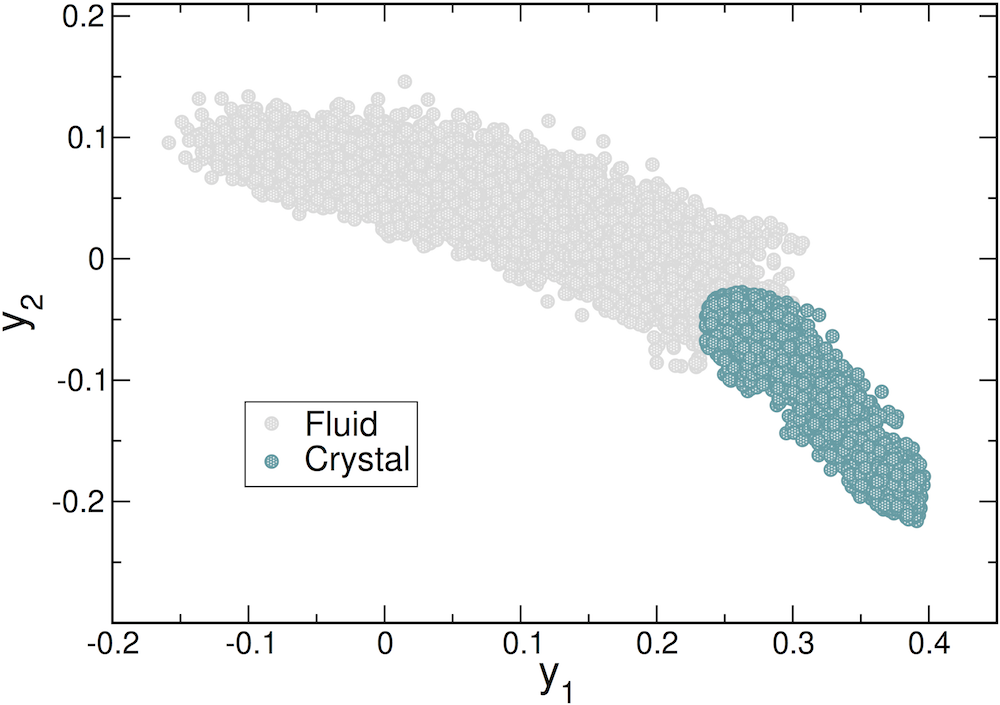}
\caption{Projection of the vectors $\mathbf{Q}(i)$ evaluated from the six snapshots in Fig. \ref{fig:snap_wca_nuc} onto the 2-dimensional space found by the encoder. Colors represent the distinct environments identified by the clustering.} 
\label{fig:scatter_wca_nuc}
\end{figure}

The two-dimensional projection found by the encoder and the results of the clustering performed in this space are shown in Fig. \ref{fig:scatter_wca_nuc}. This identification is used to colour the particles in Figure \ref{fig:snap_wca_nuc}. As expected, the  unsupervised learning method identifies two distinct particle environments, corresponding to the fluid and  crystal phases. 
% \EB{OLD} In excellent agreement with the standard classification method presented in Sec. \ref{sec:comparison}, we find a fraction of crystalline particles in the six snapshots of $2\%$ (a), $12\%$ (b), $21\%$ (c), $30\%$ (d), $64\%$ (e) and $98\%$ (f), respectively. \EB{END OLD}.  \EB{NEW} 
In order to quantitatively compare the results with the standard classification method presented in Sec. \ref{sec:comparison}, we evaluated the fraction of crystalline particles identified in the six snapshots. Note that, since we employ a soft clustering technique, this requires first to assign each particle to the cluster (crystal or fluid) with the largest membership probability. The results obtained with the two methods are presented in Tab. \ref{tab:nuc} and are in excellent agreement. 

\begin{table}[h]
\caption{Comparison between the fractions of crystalline particles identified in the six snapshots in Fig. \ref{fig:snap_wca_nuc} with our unsupervised algorithm and with the standard method presented in Sec. \ref{sec:comparison}.}
\centering
\begin{tabular}{lcccccc}
\toprule
Method & \multicolumn{6}{c}{Snapshot} \\ 
\cmidrule{2-7}
& (a) & (b) & (c) & (d) & (e) & (f)\\ 
\midrule
Unsupervised & $2\%$ & $12\%$ & $21\%$ & $30\%$ & $64\%$ & $98\%$\\
Standard        & $2\%$ & $13\%$ & $23\%$ & $31\%$ & $67\%$ & $98\%$\\
\bottomrule
\end{tabular}
\label{tab:nuc}
\end{table}

% \EB{END NEW}
% 
% From the autoencoder analysis we found that the most relevant BOPs for this system are $\bar{q}_6$ (RI$\sim 78\%$) and $\bar{q}_8$ (RI$\sim 14\%$).

One of the reasons behind performing a single analysis for the six snapshots in Fig.  \ref{fig:snap_wca_nuc}  was to include  sufficient statistics of both the fluid and crystalline environments. Our unsupervised learning method consists of a two-steps analysis: first the autoencoder finds a low-dimensional projection encoding the features with the largest variations within the input data, and then the clustering algorithm identifies distinct environments based on the density distribution of the data in this low-dimensional space. In the method presented in Sec. \ref{sec:comparison}, instead, a separation between the distinct phases is chosen based on the corresponding distributions of the relevant BOPs obtained from separate simulations. By including sufficient examples of both phases in the input dataset, the results of the unsupervised classification tend to (or at least are very similar to) those of the method in Sec. \ref{sec:comparison}, as demonstrated by the excellent agreement we found. However, what `sufficient statistics' means strongly depends on the system under study.

\begin{figure*}
\centering
\subfloat[]{%
\resizebox*{0.4\linewidth}{!}{\includegraphics{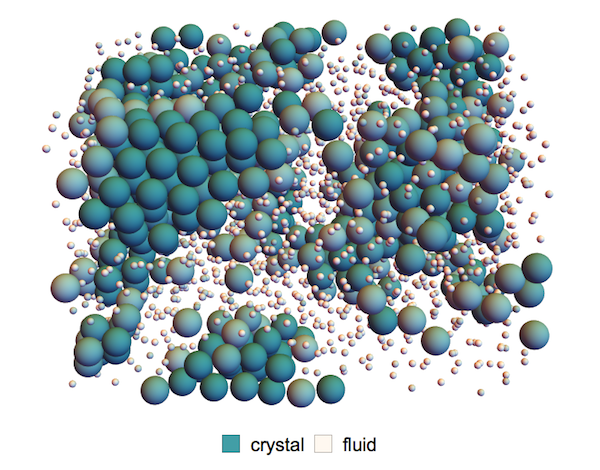}}\label{fig:snap_wca1}}\hspace{1.0cm}
\subfloat[]{%
\resizebox*{0.45\linewidth}{!}{\includegraphics{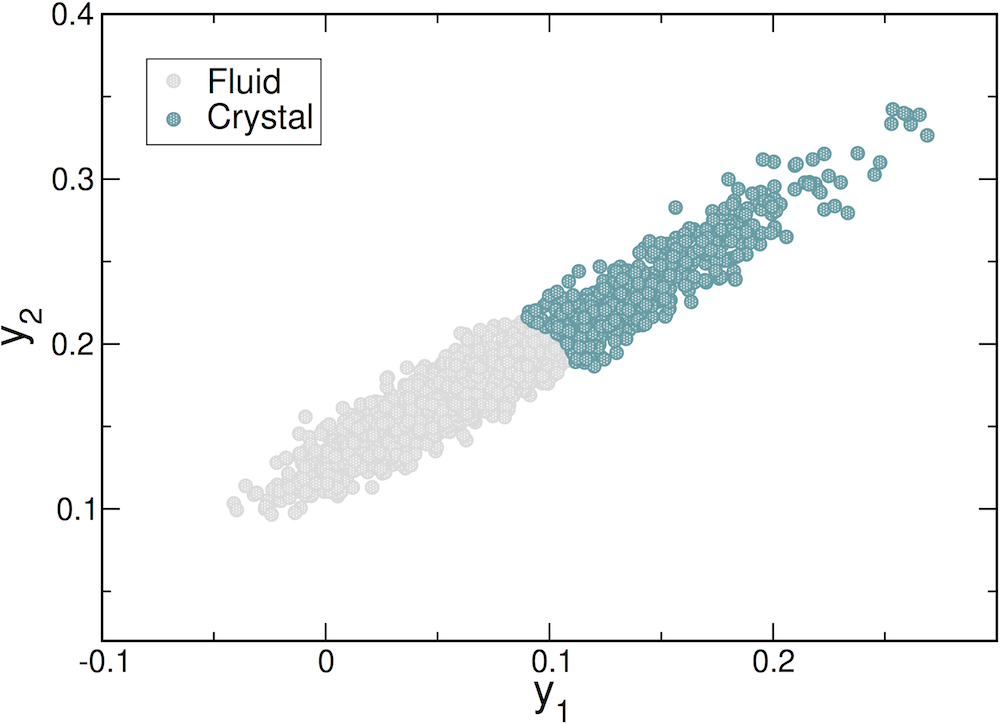}}\label{fig:scatter_wca1}}\\
\caption{Separate analysis of the snapshot in Fig.\ref{fig:snap_wca_nuc}a. (a) Classification of the snapshot under analysis. The RGB color of each particle is a linear combination of the colors of the two environments identified with the associated membership probabilities as coefficients. Fluid-like particles are displayed at $1/4$ of their actual size. (b) Projection of the vectors $\mathbf{Q}(i)$ onto the 2-dimensional space found by the encoder. Colors represent the distinct environments identified by the clustering.}\label{fig:wca1}
\end{figure*}

To obtain a better estimate of what `sufficient statistics' means for this system, we performed a separate analysis of each of the snapshots in Fig. \ref{fig:snap_wca_nuc}. For snapshots (b), (c), (d) and (e) we found very similar results to the previous analysis, while, as expected, for snapshots (a) and (f), where only a very small fraction of particles is in one of the phases, we found a different classification. As an example, we report in Fig. \ref{fig:wca1} the results obtained for snapshot (a). Specifically, in Fig. \ref{fig:scatter_wca1} we show the two-dimensional projection found by the encoder and the results of the clustering performed in this space. Note that this projection is different from the one presented in Fig. \ref{fig:scatter_wca_nuc}, as the autoencoder is trained on a different data set with different characteristics. Additionally, in Fig. \ref{fig:snap_wca1} particles are colored according to this classification. Again, the unsupervised learning identified two distinct environments, that we associate with crystalline and fluid-like particles. However, the amount of crystalline order is much larger compared to the previous classification: about $20\%$ of the particles are classified as crystalline (in the previous analysis it was only $2\%$). Interestingly, if we look closer at the snapshot in Fig. \ref{fig:snap_wca1}, we find that the particles recognized as crystalline have indeed a higher local order than in a standard disordered fluid, meaning that the unsupervised learning classification is still reasonable.  This has been seen before in nucleation studies. For instance, in Ref. \onlinecite{hsnuc}, where hard-sphere nucleation was studied at a pressure of $\beta P \sigma^3=17$, slightly different tuning of the order parameter resulted in significantly different results for the size of the largest crystalline cluster in the system  (from $\sim$ 30-100). Nonetheless, the different order parameters all predicted essentially the same nucleation rate.
% 
% is in the choice of the boundary between the two environments, which results in a stricter or looser definition of the crystalline environment. This is a common situation also with more standard order parameters. In the nucleation study in Ref. \onlinecite{hsnuc}, for instance, slightly different tuning of the order parameter gave very different results for the largest crystalline cluster in the system, without however affecting the quantitative study of the process, i.e. the estimation of nucleation rates and free energy barriers was not affected by the specific tuning of the order parameter.  

\section{Conclusions}

In summary, we have introduced a simple, fast, and easy to implement unsupervised learning algorithm for recognizing local structural motifs in colloidal systems. 
This algorithm makes use of standard BOPs to describe local environments, an autoencoder for dimensionality reduction, and GMMs for clustering the results.  
We have applied it to a wide variety of systems, ranging from simple isotropically interacting systems (hard spheres, WCA particles) to binary mixtures, and even anisotropic hard cubes.  In all cases, the algorithm performed very well, and we were able to identify local environments to a similar precision as ``standard'' - manually-tuned and system-specific - order parameters.

Moreover, we exploited the analytical mapping defined by the autoencoder to extract extra information on the systems analyzed. Specifically, in all cases we could identify the relevant symmetries underlying the main differences among the distinct particle environments found in the system. Interestingly, in the binary system we studied, this analysis revealed that the same species BOPs were the most important for distinguishing the different particle environments. Finally, in the last example on nucleation and crystal growth, we also explored the possible difficulties one can encounter when only a very small fraction of particles is in one specific environment, and we showed how including more snapshots, i.e. more statistics, could benefit the results of the analysis in such cases.

% 
% This has the advantage that it is simple and rapid to implement, but as a result we do not use all the knowledge we have available. This should be contrasted wiht the algorithm by ? where they used ...

% CNA vs Steinhaart order parameters
% 
% Diffusion map vs autoencoder
% 
% 
% 
% 
% we don't use everything we know)
% 
% 
% 
% 

\begin{acknowledgments}
We would like to thank Frank Smallenburg, Gabriele Maria Coli, and Berend van der Meer for providing us with some of the configurations we analyzed in this paper. We would also like to thank Sudeep Punnathanam, Frank Smallenburg, and Sela Samin for many useful discussions. 

L. F. and E. B. gratefully acknowledge funding from the Netherlands Organisation for Scientific Research (NWO) [grant number 16DDS004].
\end{acknowledgments}

%\nocite{*}
\bibliography{paper}% Produces the bibliography via BibTeX.

\end{document}